\documentclass[a4paper,fleqn]{cas-sc}

\usepackage[authoryear,longnamesfirst]{natbib}
\usepackage{tabularx}
\usepackage{float}
\usepackage{longtable}
\def\tsc#1{\csdef{#1}{\textsc{\lowercase{#1}}\xspace}}
\tsc{WGM}
\tsc{QE}
\tsc{EP}
\tsc{PMS}
\tsc{BEC}
\tsc{DE}

\begin{document}
\let\WriteBookmarks\relax
\def\floatpagepagefraction{1}
\def\textpagefraction{.001}

\shorttitle{IoT Digital Privacy}

\shortauthors{Shini Girija et~al.}

\title [mode = title]{Digital Privacy in IoT: Exploring Challenges, Approaches and Open Issues}                      



%
\author[1]{Shini Girija}[type=editor,
                      orcid=0000-0002-0984-5769]



\ead{p20230011@dubai.bits-pilani.ac.in}


\credit{Conceptualization of this study, Data curation, Writing - Original draft preparation, Methodology, Software}

\affiliation[1]{organization={Department of Computer Science, Birla Institute of Technology and Science, Pilani},
    addressline={Dubai Campus}, 
   city={Dubai International Academic City},
    state={Dubai},
  country={United Arab Emirates}}

\author[1]{Pranav M. Pawar} []
\cormark[1]
\ead{pranav@dubai.bits-pilani.ac.in}

  \credit{Conceptualization of this study, Data curation, Writing - Original draft preparation}
\author[1]{Raja Muthalagu} []%
\ead{raja.m@dubai.bits-pilani.ac.in}
\cormark[1]
\credit{Conceptualization of this study, Data curation, Writing - Original draft preparation}

  \author[1]{Mithun Mukherjee}
  \ead{mithun@dubai.bits-pilani.ac.in}
  \credit{Conceptualization of this study, Data curation, Writing - Original draft preparation}







\begin{abstract}
Privacy has always been a critical issue in the digital era, particularly with the increasing use of Internet of Things (IoT) devices. As the IoT continues to transform industries such as healthcare, smart cities, and home automation, it has also introduced serious challenges regarding the security of sensitive and private data. This paper examines the complex landscape of digital privacy in IoT ecosystems, highlighting the need to protect personally identifiable information (PII) of individuals and uphold their rights to digital independence. Global events, such as the COVID-19 pandemic, have accelerated the adoption of IoT, raising concerns about privacy and data protection. This paper provides an in-depth examination of digital privacy risks in the IoT domain and introduces a clear taxonomy for evaluating them using the IEEE Digital Privacy Model. The proposed framework categorizes privacy risks into five types: identity-oriented, behavioral, inference, data manipulation, and regulatory risks. We review existing digital privacy solutions, including encryption technologies, blockchain, federated learning, differential privacy, reinforcement learning, AI, and dynamic consent mechanisms, to mitigate these risks. We also highlight how these privacy-enhancing technologies (PETs) help with data confidentiality, access control, and trust management. Additionally, this study presents AURA-IoT, a futuristic framework that tackles AI-driven privacy risks through a multi-layered structure. AURA-IoT integrates adversarial robustness, explainability, transparency, fairness, compliance, dynamic consent, and policy enforcement mechanisms to ensure digital privacy, security, and accountable IoT operations. Finally, we discuss ongoing challenges and potential research directions for integrating AI and encryption-based privacy solutions to achieve comprehensive digital privacy in future IoT systems.
\end{abstract}


\begin{highlights}

\item Proposed a risk-driven taxonomy of digital privacy risks in IoT devices.
\item Presented existing digital privacy solutions to mitigate these risks.
\item Developed AURA-IoT that connects digital privacy risks, solution types, and evaluative criteria.

\end{highlights}


\maketitle

\section{Introduction}

The significance of digital privacy in the modern digital landscape cannot be emphasized \cite{adams2023meaning}. Since 2020, amid the challenges posed by the COVID-19 pandemic, there has been a discernible increase in data breaches and privacy violations, making digital privacy incidents a primary concern \cite{majeed2021comprehensive}. By 2020, a significant turning point had been reached, with 21\% of consumers experiencing online account hacking and 11\% of people experiencing data theft \cite{srinivas2022being}. Over time, cyber-attacks that affected 12\% of Internet users brought attention to the risks of sharing personal information online \cite{Statista2023}. Data breaches are becoming the "absolute norm" in the digital era, ranging from minor attacks involving ransomware to major breaches that impact millions of people across industries such as government agencies, healthcare, banking, and retail. Privacy protection is more critical than ever as technology continues to permeate every aspect of daily life, from social media to online commerce. Digital privacy protects people's liberty, security, and well-being by granting them the fundamental right to manage who can access their personal information in digital settings \cite{fainmesser2023digital}. Personal information has grown in value with the advent of Artificial Intelligence (AI), social media, e-commerce, and online banking. These technologies are convenient, but they also pose new security and privacy risks. Maintaining the confidentiality, availability, and integrity of personal data in this era of interconnection requires an awareness of the implications and challenges of digital privacy. Furthermore, the significance of digital privacy increases with the adoption of the Internet of Things (IoT), which connects usual gadgets to the internet and gathers data \cite{salam2024internet} \cite{bartlett2020privacy}. Large volumes of data are generated by interactions with IoT devices, underscoring the need for robust privacy safeguards to prevent misuse and unauthorized access to personal data. 

\begin{figure*}[!t]
\centering
\includegraphics[width=4in]{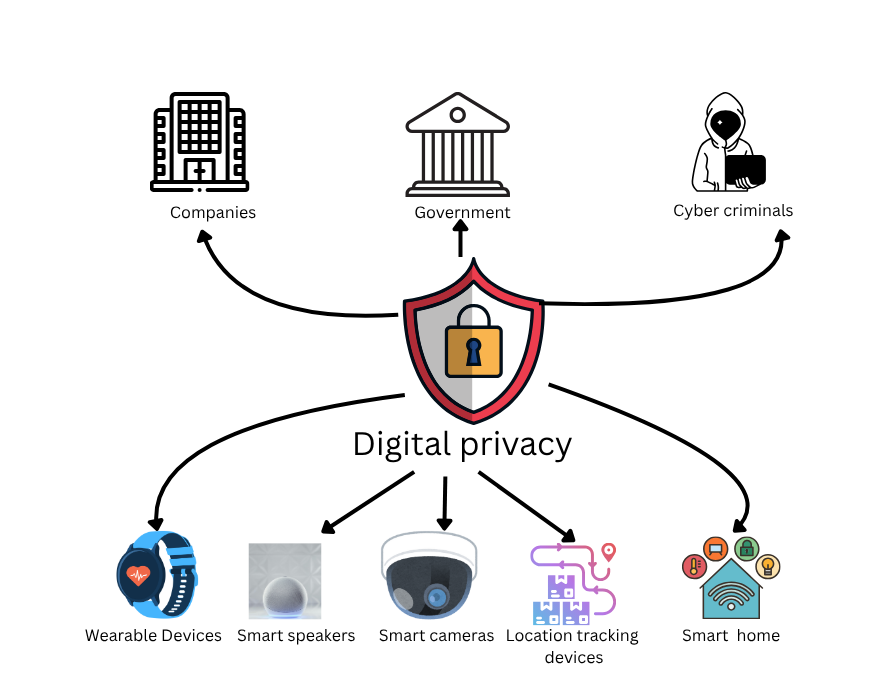}
\caption{IoT devices in digital privacy domain}
\label{fig1}
\end{figure*}

Companies, the government, and cybercriminals are the three main entities that gather and utilize personal data from IoT devices, such as fitness trackers, smart home assistants, and location-tracking devices. First, companies offer a wide range of free apps and services, but they profit from the data collected from users. Second, although government surveillance technology is intended to improve security without imposing regulations, it may infringe upon individual liberties. Third, cyber criminals sell personal information on the dark web through malware and phishing. Fig. \ref{fig1} presents the IoT devices in the digital privacy domain.

\subsection{Motivation}

The necessity to protect personal data has become increasingly important to internet users, governments, tech businesses, and online advertisers due to the development of generative AI and the growing role of IoT devices in people's daily lives \cite{huang2023security}. From wearable fitness trackers to smart home appliances, IoT devices have become an integral part of our lives due to their inherent internet connectivity and data transfer capabilities. These gadgets make life easier, but they also present serious risks to digital privacy \cite{singhai2023investigation} \cite{najmi2023survey}. IoT devices constantly transmit data, raising concerns about security and privacy protection. The likelihood of privacy violations and unauthorized access to sensitive data increases with the number of connected devices. As a result, in the IoT ecosystem, businesses and consumers must prioritize digital privacy. Meneghello et al. (2019) \cite{meneghello2019iot}  highlighted a pressing demand for automated techniques that could facilitate the early identification of privacy concerns in IoT systems and provocatively defined IoT as the Internet of Threats, highlighting the pervasiveness of the privacy problem in practice.

\begin{figure*}[!t]
\centering
\includegraphics[width=5in]{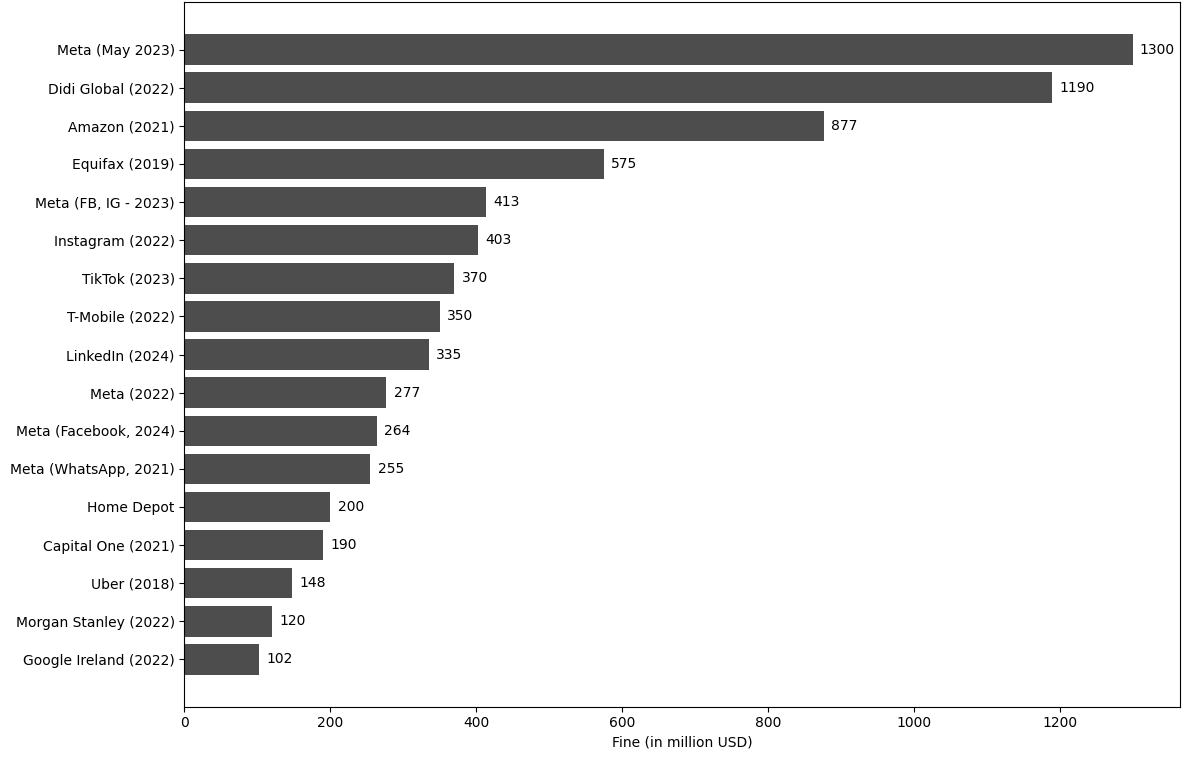}
\caption{Largest data privacy violation fines \cite{statista2025privacyfines}}
\label{fig2}
\end{figure*}

Millions of Facebook users' personal information was gathered in the 2010s by the British consulting firm Cambridge Analytica for political advertising purposes without their knowledge or agreement to sway political campaigns \cite{matara2024facebook}. People worldwide are now more aware of the concerns associated with digital privacy after Facebook owner Meta paid a \$725 million (£600 million) settlement to resolve a lawsuit over this. Fig. \ref{fig2} shows the most significant data privacy violation fines, penalties, and settlements worldwide as of January 2025. Fig. \ref{fig2} makes it evident that penalties for data privacy violations are on the rise, underscoring the necessity of paying close attention to identifying various digital privacy threats and the need for effective digital privacy policies to be enforced and addressed.

Researchers and practitioners are focusing on various novel strategies to protect digital privacy in response to the growing privacy challenges in IoT systems. We outline the key approaches as:-
\begin{itemize}
    \item Encryption: In IoT systems, lightweight encryption methods provide effective and safe communication, protecting digital privacy and preserving maximum resource efficiency \cite{medileh2020flexible} \cite{abdulraheem2020efficient}.
    \item Machine Learning (ML) and Deep Learning (DL)-based: ML and DL-based approaches are being utilized increasingly to enhance digital privacy in IoT systems by intelligently anonymizing data, detecting threats, and controlling access. Federated learning (FL) and reinforcement learning (RL) are ML/DL approaches utilized to protect digital privacy by facilitating decentralized data processing and adaptive privacy management without jeopardizing the security of individual data. FL is a distributed collaborative technique that enables data training across various devices with a central server, eliminating the need to share actual datasets \cite{nguyen2021federated} \cite{li2021survey}. It is intended to be used to develop intelligent and privacy-enhanced IoT systems. RL leverages adaptive algorithms to dynamically optimize privacy settings based on real-time data privacy hazards and user preferences \cite{qin2021privacy} \cite{shateri2020privacy}. 
    \item Artificial Intelligence: AI techniques focus on creating supervised and unsupervised procedures, meta-heuristics, or reasoning techniques to identify possible privacy breaches or protect privacy in IoT systems \cite{giordano2022use} \cite{choudhuri2023privacy}. AI systems can analyze data, identify normal behaviors, and detect security risks or privacy loss. They can respond promptly to potential threats, aid in forecasting, and help create data laws to reduce legal issues \cite{devineni2024ai} \cite{selvakumar2025balancing}. This allows businesses maintain data privacy and security within legal boundaries.
     \item Differential Privacy: Differential privacy (DP) has become popular for protecting data since it makes it possible to create stringent mathematical privacy guarantees \cite{demelius2025recent}. DP offers a strong foundation for safeguarding private information while preserving its computational utility. Essentially, a differentially private algorithm accepts the data of several parties as input and outputs a result that reveals very little about any of the parties \cite{angrisani2025quantum}. DP measures how well individual privacy is preserved in IoT systems while allowing for the noisy disturbance of data to expose valuable aggregate information \cite{husnoo2021differential} \cite{zhao2020local}.
    \item Consent mechanisms: Digital privacy consent approaches include dynamic consent, which enables real-time adjustments to consent preferences for continuous data control, and granular consent management, which gives consumers comprehensive control over data sharing \cite{pathmabandu2023privacy} \cite{wakenshaw2018mechanisms}.
\end{itemize}

 \subsection{Related Works}\label{section 2}

\begin{table}[h!]
\centering
\caption{Comparison of existing surveys on digital privacy}\label{tab:table1}

\small
\renewcommand{\arraystretch}{1.2} 
\setlength{\tabcolsep}{4pt}       
\begin{tabular}{p{2cm} p{2.5cm} p{3cm} p{4.5cm}p{4cm}}
\hline
\textbf{Category} & \textbf{Refs} & \textbf{Focus} & \textbf{Method} & \textbf{Limitations} \\
\hline
Privacy risks in information systems 
& Rath \& Kumar, 2021~\cite{rath2021information} 
& Privacy at individual, collective, institutional, and social levels, considering contextual, cultural, and legal factors 
& Literature review on IS privacy issues 
& Limited demographic diversity and real-world applications \\
& Saura et al., 2021~\cite{saura2022evaluating} 
& Analyzing security and privacy risks in social network-based information systems 
& Exploratory data-driven approach using sentiment analysis on tweets 
& Only Twitter data considered \\
Privacy risks in crowdsourcing applications
& Kumar \& Faisal, 2024~\cite{kumar2024comprehensive} 
& Privacy in crowdsourcing applications 
& Framework analysis 
& Ignores the dynamic nature of crowdsensing privacy \\

& Shahzad, 2024~\cite{shahzad2024privacy} 
& Privacy and security concerns in online crowdsourcing platforms 
& Systematic literature review applying Information Privacy Theory and Technology Acceptance Model 
& Limited empirical validation \\
Privacy risks in education
& Paul \& Knor, 2022~\cite{paul2022taxonomy} 
& Privacy education: domain, delivery method, mode, and subject 
& Domain-based classification 
& Overlooks cultural and geographic variations \\

& Kamenskih, 2022~\cite{kamenskih2022analysis} 
& Security and privacy challenges in integrated smart educational environments 
& Analytical review of key technologies and user interaction points within smart education systems 
& Lack of empirical evaluation \\

Privacy risks in data analytics applications
& Kasera et al., 2023~\cite{kasera2023right} 
& Security measures, challenges, and privacy solutions 
& Privacy-enhancing methods in literature 
& Lacks focus on scalability and real-world systems \\

& Malina et al., 2021~\cite{malina2021post} 
& IoT privacy issues across device, network, data, ethics 
& Research on PET systems, IoV case study 
& Misses emerging IoT threats \\

Privacy risks in healthcare applications
& Mahadik et al., 2024~\cite{mahadik2024digital} 
& Privacy in healthcare 
& AI-integrated privacy analysis 
& Limited empirical validation \\

& Madhu, 2025~\cite{madhu2025survey} 
& Privacy and security threats in IoT-enabled healthcare systems 
& Analytical discussion of real-world vulnerabilities, potential threats, and current mitigation strategies in healthcare IoT 
& Lack of empirical evidence and real-case studies \\
\bottomrule
\end{tabular}
\end{table}

 There are several papers that examine digital privacy, but its use in IoT applications is not  much explored, indicating that research is still in its infancy. We categorize the papers based on IoT applications as shown in Table \ref{tab:table1} follows:- 

 \begin{itemize}
     \item \textit{Privacy risks in information systems:}-  Rath et al. \cite{rath2021information} examined a wide range of research on privacy concerns in information systems, emphasizing essential topics, including societal, organizational, and cultural differences in privacy attitudes and regulations. However, the study's scope is constrained because it only draws on a few studies and doesn't thoroughly examine multiple domains, multilevel analyses, or practical applications beyond personal privacy issues.

     \item \textit{Privacy risks in crowdsourcing applications :-} Kumar and Faizal \cite{kumar2024comprehensive}  categorized privacy risks in crowdsourcing applications according to current anonymity issues, and described solutions like blockchain and encryption. However, the study did not cover real-world validation and the dynamic, changing nature of crowdsensing applications.
     \item \textit{Privacy risks in education:-}  Paul and Knox \cite{paul2022taxonomy} classified digital privacy-based education research according to the application domain, teaching delivery method, teaching mode (e.g., shared, synchronous, asynchronous, experiential), and data subject (personal or third party). The limitation of the research is its limited focus on cultural and geographical differences in privacy education.
     \item  \textit{Privacy risks in Data analytics applications:-} Kasera et al. \cite{kasera2023right} categorized concerns in the significant digital data privacy domain according to security measures, privacy challenges, and the characteristics of data analytics and blockchain technologies. This study pays little attention to the scalability and practical application of privacy-preserving systems. Malina et al. \cite{malina2021post} categorized IoT security and privacy issues into device-level, network, data, and ethical factors, emphasizing AI-driven solutions, post-quantum cryptography, and privacy-enhancing technologies. However, the study failed to consider other emerging IoT security threats and methods. Aqeel et al. \cite{aqeel2022review} classified IoT security risks into several categories, including network, access-level, active, passive, strategy-level, physical, logical, hardware-compromised, software, firmware, and cryptanalysis attacks. The research's shortcomings include its lack of attention to the scalability and practical application of innovative solutions like blockchain and artificial intelligence for IoT security.
     \item  \textit{Privacy risks in healthcare applications:-}  Mahadik et al. \cite{mahadik2024digital} focused on privacy laws and regulations while classifying digital privacy concerns in healthcare into risk assessment methodologies, data encryption techniques, access control mechanisms, and the use of AI. One drawback would be that the survey only examines digital privacy concerns in specific IoT applications in the healthcare sector rather than exploring the broader range of IoT applications in different industries. This could restrict the findings' applicability to other IoT use cases.
 \end{itemize}
\subsection{Novelty and Research Contributions}

Although prior works have reviewed IoT privacy and security challenges, a comprehensive examination of digital privacy concerns in IoT devices and applications is needed. Existing surveys often focus on specific IoT applications, providing limited attention to privacy preservation strategies and offering only partial coverage of the broader privacy landscape. This survey bridges these gaps by integrating conceptual risks, technical mitigation strategies, and architectural design.

The key contributions and novel aspects of this study include:

\begin{itemize}
    \item \textbf{Risk-Driven Taxonomy}: Proposing a novel digital privacy risk categorization framework with five groups: identity-oriented risks, behavioral risks, inference risks, data manipulation risks, and regulatory risks. This taxonomy semantically maps privacy concerns to IoT system layers and phases of the security lifecycle.
    
    \item \textbf{Comprehensive Discussion of Privacy Solutions}: Discussing digital privacy solutions that address these privacy risks in IoT, integrating both traditional and emerging approaches. Section 3 presents the existing digital privacy solutions in IoT.
    
    \item \textbf{Integration of AI and PETs}: A comparative analysis of AI-driven privacy-enhancing technologies (PETs), including federated learning, differential privacy, encryption, blockchain, and consent, examines the complementarity and interdependence relationships along with the trade-off axes. Section 4 illustrates the role of AI in the digital privacy of IoT.
    
    \item \textbf{Architectural Contextualisation}: Connects the operational flow across the IoT layers with privacy goals of fairness, explainability, and compliance, introducing the AURA-IoT conceptual framework.
    
    \item \textbf{Bridging Technical and Ethical Dimensions}: Coordination of digital privacy in IoT and policy enforcement, user consent, AI explainability, and traditional technical barriers provides a rich and ethical perspective on the fractured nature of digital privacy.
    
    \item \textbf{Exploration of Future Directions}: Exploring future research directions in IoT digital privacy. Section 6 discusses the future research scopes in IoT digital privacy.
    
\end{itemize}

These contributions provide a deeper analysis and an integrated approach that extends beyond existing IoT privacy research surveys, paving a clear foundation for emerging AI-driven digital privacy research efforts in the IoT domain.

The rest of this survey is organized as follows.  Section \ref{section3} presents the proposed digital privacy risk categorization, and Section \ref{section 4} presents the existing digital privacy solutions in IoT. Section \ref{section 5} highlights the role of AI in the digital privacy of IoT.  Section \ref{section5} presents the proposed AURA framework for ensuring digital privacy in IoT to address AI-related privacy risks. Section \ref{section6} future research scopes in IoT digital privacy. Finally, Section \ref{con} concludes the paper.
\section{Digital Privacy risks categorization} \label{section3}

 \begin{figure*}[!t]
\centering
\includegraphics[width=6in]{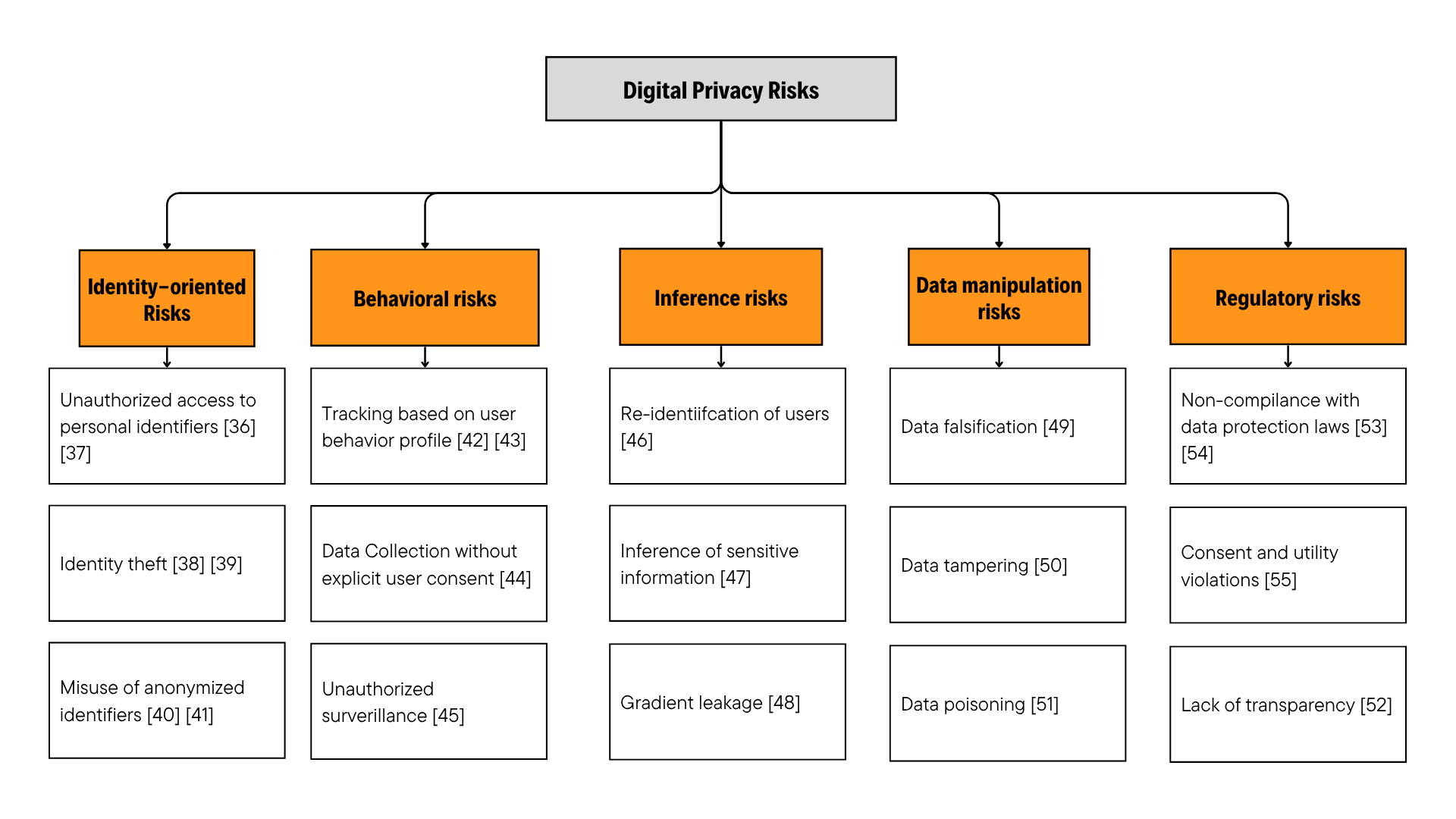}
\caption{Taxonomy of digital privacy risks}
\label{fig3}
\end{figure*}

The IEEE Digital Privacy Model (DPM) proposed a secure framework for digital privacy with clear expectations and significant components \cite{ieee_digital_privacy_committee}. Regardless of location, the method focuses on the user, their privacy expectations, and the elements that impact online privacy. This evolving model demonstrates how technical, regulatory, economic, legislative, legal, individual, societal, and cultural influences on people's expectations of privacy are more achievable by using confidentiality and integrity as well as access and observability of people's identities, behaviors, inferences, and transactions in any digital ecosystem to represent digital privacy for individuals \cite{ieee_digital_privacy_article}. The primary goal of this research is to establish a privacy risk framework for assessing privacy risks associated with the IoT domain based on the characteristics of privacy outlined in the DPM model. Identity-oriented, behavioral, inference, data manipulation, and regulatory risks comprise the five categories that make up our proposed risk framework. The taxonomy of the identified digital privacy risks and the main influencing factors are shown in Fig. \ref{fig3}.

The subsequent subsections provide more details on these risk categories.
IoT applications vary in their needs and limitations, and the degree of privacy protection required may change based on the application's legal and ethical considerations, the utility of sensitive data, the sensitivity of the data being utilized, and the possible consequences of a privacy breach. For example, a law enforcement program with public security concerns would have different privacy needs than a smart healthcare system, which handles sensitive patient data. While privacy must be balanced with the need to ensure public safety in law enforcement, privacy is a critical ethical and legal necessity that must be upheld in smart healthcare. These characteristics are taken into account in the proposed privacy risk framework.

\subsection{Identity-Orientated Risks}
Informational identity includes all forms of information about a person, such as digital representations of identity (such as legal representations in addition to online user accounts, social networks, preferences and attributes inferred from usage or behavioral data), legal representations of identity (such as driver's license, passport, or national identity card, contracts), and any other information describing the person \cite{giannopoulou2023digital}. Access control and identification technologies offer the necessary framework for connecting data between users' devices with distinct identities and delivering streamlined, integrated services. However, profiling techniques based on linked records may disclose sensitive information about a user's identity and personal life, thereby violating their right to privacy and potentially leading to social, economic, and other types of discrimination \cite{christen2020linking}.

Hackers have frequently exploited their access to IoT devices to steal personal data for financial gain or identity theft \cite{ahvanooey2021modern}. In other instances, they have tampered with industrial equipment or transportation networks, such as cars, trains, and aeroplanes, causing physical harm. 
Additionally, an attacker can obtain unauthorized access to secured network resources by assuming or imitating another user's identity. An attacker will frequently attempt to pose as a senior executive to get around security measures or compromise corporate information \cite{shafiq2022rise}. In other instances, hackers have attempted to deceive victims into installing malicious software by disguising well-known public figures—such as celebrities—as benign items, like photo attachments.

Misusing anonymized, or privacy-preserving, identifiers poses serious identity-oriented risks to IoT privacy. Privacy-preserving identifiers are information containers used to identify or authorize a device or a user without necessarily disclosing the device holders' identities or other personal information \cite{akil2020privacy}. IoT generates vast amounts of data, which must be anonymized before being shared with the public or other parties to reduce the risk of re-identification and protect sensitive data. Information that can be used to identify specific people can be eliminated through data anonymization processes \cite{yang2023attack}. However, the misuse of privacy-preserving identifiers and improper data anonymization can increase the likelihood of re-identification, leading to privacy breaches. 

\subsection{Behavioral Risks}

User behavior-based tracking and profiling pose a serious privacy risk. A user profile includes location, recent activity, nearby items, and other users, as well as social context. A user's social networks comprise their social connections with other users, representing their social environment. The collection of user data to identify their interests through correlation with profiles and data from different sources carries a risk \cite{ogonji2020survey}. Profiling may result in privacy violations if the information is utilized for pricing discrimination, social engineering, or unsolicited advertisements \cite{rizi2022systematic}. When developing and evaluating data profiles, the difficulty is in striking a balance between the user's interests and privacy constraints. Obtaining and selling user profiles on the data marketplace without the individual's consent is also regarded as a violation of their privacy.

Additionally, continuous IoT device monitoring without the user's explicit consent may result in the unintentional recording of sensitive or private moments, compromising individual privacy. For example, smart home assistants like Alexa may inadvertently record private conversations \cite{edu2020smart}, and an attacker within or close to a smart home environment may be able to take advantage of the wireless medium that these devices are built on to steal sensitive information about users and their activities from the encrypted payload (i.e., sensor data), thus violating user privacy.

Behavioral profiling involves collecting long-term data about a user's activities and using that knowledge to customize the user interface with minimal user consent \cite{karale2021challenges}. For example, some smart healthcare use cases, such as fitness trackers, disease tracking, and emotion detection, have been shown to track, gather, monitor, and retain user data without the user's explicit consent.

\subsection{Inference Risks}

Inference concerns may still reveal private information about the raw data from local models or gradients \cite{kapoor2024computation}. An adversary with malicious intent or an honest but inquisitive application server may use inference attacks to reveal the participants' personal information, including their routines, habits, and places of employment and residence. This information leakage also applies to anonymous donations, as location traces can deduce participants' identities, daily schedules, and frequently travelled routes based on publicly accessible data \cite{chen2024federated}. For instance, applications that capture audio using a mobile phone's microphone have the potential to act like smart spies and be utilized by adversaries to ascertain the context. Therefore, it is imperative to withstand these attacks to accomplish computing that completely preserves privacy.

Adversaries extract critical information from the machine learning model through data leakage at the central server, resulting in the loss of data items \cite{singla2024privacy}. Sensitive information leaks during data transmission must be detected using an appropriate method. Devices with default permissions in the IoT setting frequently leak data. Most privacy issues are caused by the disclosure of such information, including roughly 29\% of unintentional leakage of private or sensitive information, 16\% of intellectual property materials, and 15\% of financial data.

\subsection{Data manipulation Risks}

IoT devices are prone to data manipulation, which allows unauthorised parties to change data and produce incorrect information and possible harm \cite{lee2021improving}. Data collection, transmission, and storage are only a few of the steps at which this manipulation may take place. Data falsification is the deliberate change of data to mislead or deceive, much like manipulation risks \cite{singh2022framework}. Data manipulation can facilitate information linkage and user profiling, which in turn compromises privacy by combining data from multiple sources to generate comprehensive profiles of individuals. Additionally, the attacker tampers with the intercepted information and sends it to the recipient, where altering or changing the content may have serious privacy violations. Decisions based on manipulated information may result in unfair, discriminating decisions regarding employment, insurance, or other services. Additionally, through manipulating sensitive user information, an adversary might initiate poisoning attacks that result in erroneous IoT device operation and privacy violations \cite{huang2022defending}. For instance, the attacker can force the system to make targeted misclassifications or lower the model classification accuracy by introducing specially crafted poisons into the training dataset of the DNN model in the traffic sign recognition system.
\begin{figure*}[!t]
\centering
\includegraphics[width=4in]{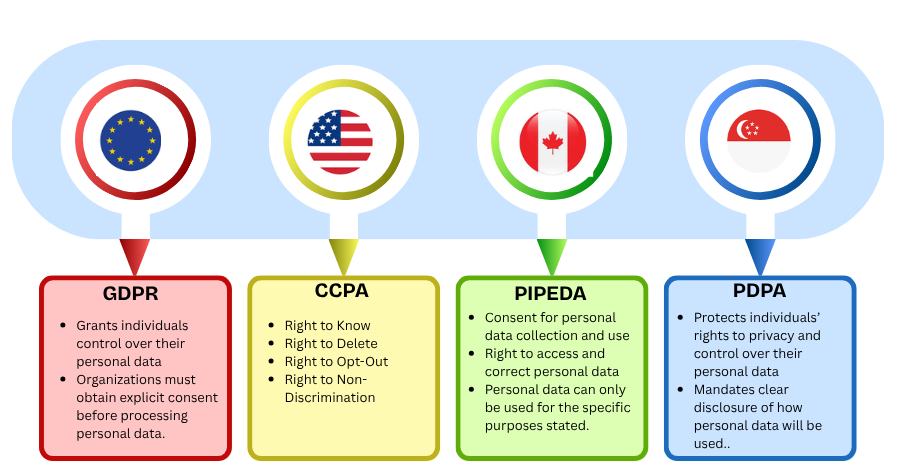}
\caption{Digital privacy laws}
\label{fig4}
\end{figure*}
\subsection{Regulatory Risks}
The constant data collection by IoT devices, often without users' knowledge or consent, poses a risk of illegal surveillance and regulatory issues. The scope and intent of data gathering by IoT devices are frequently unknown to users, making it more challenging for them to comply with open consent and transparency regulations \cite{cook2023security}. 

Organization laws as shown in Fig. \ref{fig4}, including the General Data Protection Regulation (GDPR) \cite{munk2021internet}, California Consumer Privacy Act (CCPA) \cite{aljeraisy2021privacy}, Personal Information Protection and Electronic Documents Act (PIPEDA) \cite{aljeraisy2021privacy} and Personal Data Protection Act (PDPA) \cite{ghani2020overview}, have offered guidelines and regulations on how to guarantee the protection of personal information. The diversified and pervasive nature of IoT devices makes it difficult to ensure compliance with international data protection laws, such as GDPR. Nevertheless, these rules frequently function independently, resulting in uneven adherence throughout the extensive IoT network. Inadequate efforts to adhere to privacy laws and policy guidelines can compromise privacy and have a serious impact on organizations. First, most privacy regulations severely penalize companies for violating privacy, which may result in financial loss. Furthermore, an organization's ability to compete in the market may be hampered by non-compliance with legislation, as it may make them appear untrustworthy to rivals. Lastly, reputation damage may arise from non-compliance with privacy laws.

The classification of digital privacy risks in IoT systems reflects the current challenges associated with securing sensitive personal information in interconnected networks. IoT digital privacy concerns encompass complex issues such as data manipulation, inference attacks, behavioral profiling, identity theft, and regulatory non-compliance. Additionally, the dynamic environment of IoT, with its unparalleled data flow patterns and novel sources of attacks, makes data protection even more challenging. Combining strong encryption and privacy-protecting AI, dynamic consent policy systems, and following regulatory policies are necessary to tackle these issues. These systems will benefit from an updated approach to risk analysis and mitigation if the objective is to maintain the security, visibility, and coherence of IoT systems with users’ desires for privacy and digital interaction.
\begin{table}[h!]
\centering
\caption{Mapping of Privacy Risk Types to IoT System Stack and Security Lifecycle Phases}
\label{tab:risk_lifecycle}
\small
\renewcommand{\arraystretch}{1.2} 
\setlength{\tabcolsep}{4pt}       
\begin{tabular}{p{2.5cm} p{2.5cm} p{3cm} p{4.5cm}}
\hline
\textbf{Risk Type} & \textbf{IoT Stack Layer} & \textbf{Security Lifecycle Phase} & \textbf{Mitigation Mechanisms} \\
\hline
Identity-oriented \cite{yao2021security}& Perception / Edge & Data Collection & Localized identification, encryption, and access control \\
Behavioral \cite{mustafa2021iot} & Network / Application & Data Transmission / Access & Dynamic consent enforcement and privacy preference management \\
Inference \cite{duan2022distributed} & Cloud / AI  & Data Processing & Federated learning and differential privacy for privacy-preserving analytics \\
Data Manipulation \cite{ullah2021secure} & Fog / Edge & Data Storage / Aggregation & Integrity validation, adversarial robustness, and secure updates \\
Regulatory \cite{shahi2024navigating} & All Layers & Cross-lifecycle & Policy enforcement, auditability, and compliance monitoring \\
\hline
\end{tabular}
\end{table}

\subsection{Operational Mapping of Privacy Risks to IoT Architecture and Security Lifecycle}
This section combines the IoT architectural viewpoint with the stages of the security lifecycle to support the operational basis of the proposed five-type privacy risk classification.

Each risk category corresponds to a specific IoT system layer (perception, network, edge/fog, cloud, and application) and lifecycle phase (data collection, transmission, processing, storage, and access). This dual mapping helps clarify where privacy risks arise and how to reduce them using suitable PETs. 
Identity-related risks mainly occur in the perception and collection phases \cite{yao2021security}. Device identifiers or credentials can expose users to tracking or impersonation threats during these stages. Behavioral risks appear in the transmission and access phases, when continuous monitoring or contextual logging can reveal user habits or activity patterns \cite{mustafa2021iot}. These risks can be mitigated by using dynamic consent methods and localized identification, which enable users to maintain detailed control over their shared data. Inference risks are concentrated in the cloud and AI layers, emerging during data processing and analysis \cite{duan2022distributed}. The relation among data attributes in these stages can lead to unintended leakage of sensitive information. Current PETs, such as FL and differential privacy, mitigate these risks by reducing direct data exposure while providing data for analysis. Data manipulation risks happen during the storage and aggregation processes in the fog and edge layers \cite{ullah2021secure}. These can be minimized by implementing integrity validation, secure update procedures, and robustness checks against attacks. Regulatory risks apply to all layers and lifecycle phases, which require consistent policy enforcement, proper auditing, and regular compliance with dynamic data protection standards \cite{shahi2024navigating}. This operational mapping transforms the semantic privacy risk classification into a more practical and verifiable model that aligns with IoT architectures and security lifecycle phases, thereby strengthening its utility for real-world systems.

\section{The state of the art in IoT digital privacy} \label{section 4}
The confidentiality, integrity, and availability of user data across IoT devices must be established and maintained to guarantee the digital privacy of IoT systems. Therefore, this section will examine specific strategies for data encryption, differential privacy, and consent mechanisms that enhance the privacy of IoT devices. Data encryption techniques help protect private user data, even if it falls into the hands of unauthorized individuals. A well-known privacy innovation, differential privacy, restricts the conclusions that may be drawn about one individual from the data of another, thereby improving privacy protection, particularly against algorithmic predictions. Numerous consent mechanisms \cite{pinto2024towards} \cite{merlec2021smart} are available to protect privacy while prioritizing users' consent to disclose data.

\begin{table}[htbp]
  \caption{Review of the most recent state-of-the-art data encryption methods.}
  \label{tab:table2}
  \small
  \begin{tabular}{@{}p{2cm}p{2.5cm}p{2.5cm}p{2.2cm}p{2.2cm}p{3cm}@{}}
    \toprule
    \textbf{Refs} & \textbf{Encryption Methods} & \textbf{Advantages} & \textbf{Data Storage Domain} & \textbf{IoT Applications} & \textbf{Security Triad} \\
    \midrule
    Li et al., 2019~\cite{li2019searchable} & Searchable encryption scheme & Handles all content formats & Cloud server & Speech/image recognition & Confidentiality \\
    
    Latif et al., 2020~\cite{abd2020controlled} & Quantum walks & Lowering computing power & Cloud server & Healthcare systems & Confidentiality \\
    
    Yu et al., 2021~\cite{yu2021blockchain} & Shamir threshold cryptography & Ensure privacy against third-party implications & Blockchain & IIoT & Confidentiality, Integrity, Availability \\
    
    Singh et al., 2021~\cite{singh2021blockchain} & Homomorphic encryption & Minimal computing overhead & Blockchain & Smart grid & Confidentiality, Integrity, Availability \\
    
    Alluhaidan \& Prabu, 2023~\cite{alluhaidan2023end} & Lightweight cryptographic schemes & Lightweight, safe, and energy-efficient & Cloud server & Resource-constrained IoT applications & Confidentiality, Integrity, Availability \\
    \bottomrule
  \end{tabular}
\end{table}
\subsection{Data Encryption}

Cloud servers, which can provide enormous storage and cloud computing services, will receive the sensing data generated by sensors and devices within the IoT ecosystem \cite{li2019searchable}. When storing a large amount of encrypted data on cloud servers, the primary issue is data confidentiality, which effective and secure encryption techniques can guarantee \cite{chen2025end}. Encryption is essential for safeguarding private information in the IoT by preventing breaches and unwanted access. Methods such as stream ciphers \cite{li2019searchable}, Advanced Encryption Standard (AES) \cite{yu2021blockchain}, Elliptical Curve Cryptography (ECC) \cite{adere2022blockchain}, lightweight algorithms \cite{singh2021blockchain}, homomorphic encryption \cite{yu2021blockchain}, Attribute-based Encryption (ABE) \cite{abdaoui2021fuzzy}, and neural cryptosystems \cite{sun2021lightweight} provide a number of advantages that are specifically suited to the particular limitations and needs of IoT devices. The particular use case, device features, and security requirements all play a role in choosing the best encryption technique. However, because the cloud server cannot decrypt the encrypted data, it cannot respond to the data user's request to retrieve data containing a particular keyword. Authors 
\cite{li2019searchable} addressed the privacy protection of big data during the data storage phase and proposed a searchable encryption technique that satisfies individual privacy requirements. All file kinds, including text, audio, images, videos, and more, can be used with the proposed approach, which balances protecting privacy with guaranteeing data utility while accommodating varying privacy requirements. Another study proposed STCChain, a Shamir threshold cryptography technique for blockchain-based IIoT data safety to solve the security and privacy implications of third-party organizations \cite{yu2021blockchain}. In particular, this method encrypts the data uploaded by the IoT device and stores it in the cloud using a symmetric key. The edge gateway generates a private key to secure the symmetric key. A Shamir secret-sharing technique is used to split the private key, encrypt it, and post it on the blockchain to guard against privacy breaches and key loss. Table \ref{tab:table2} summarizes the most recent data encryption methods in terms of the encryption type,  the advantages of the proposed approach, whether the proposed solution utilizes cloud computing or blockchain technology, and the applications, in addition to the security triad achieved.

The increasing use of IoT devices and sensors by healthcare systems to collect medical data is driving the need for enhanced processing power in privacy management. However, when these technologies are sent or exchanged over networks, they still present security and privacy risks. New security measures must be created to protect the privacy of medical data and overcome these obstacles. Quantum technology developments have the potential to significantly alter cryptographic processes, enhancing data security and privacy while reducing computing power requirements. The study \cite{abd2020controlled} developed a framework for safe privacy preservation in IoT-based healthcare using quantum walks, in which patients encrypt private images, upload them to a cloud-based IoT system, and have medical professionals decode the data. The massive amount of data IoT devices create and save in cloud storage presents serious security concerns for smart grid IoT systems. 

Cyberattacks have the potential to cause significant harm to the electrical system, resulting in monetary and other damages. This study \cite{singh2021blockchain} proposes a privacy-preserving data aggregation strategy based on homomorphic encryption to provide secure data aggregation with minimal computational overhead and mitigate the detrimental effects of flash workload on prediction models. The data aggregation framework, based on blockchain and cloud computing, is designed to enhance security. Another study proposed a lightweight, safe, and energy-efficient cryptography method for IoT devices \cite{alluhaidan2023end}. It utilized a symmetrical encryption key block, a modified Feistel architecture, and a unique proxy network (SP).  The Feistel cipher sequences form the foundation for the symmetric key encryption system, which utilizes genetic algorithm principles to generate multiple rounds and sub-keys. This proposed approach offers adequate security while reducing processing cycles.

\subsection{Differential privacy}

\begin{table}[htbp]
  \caption{Review of the most recent state-of-the-art differential privacy methods.}
  \label{tab:table3}
  \small
  \begin{tabular}{@{}p{1.5cm}p{3.2cm}p{2.5cm}p{4.2cm}p{3cm}@{}}
    \toprule
    \textbf{Refs} & \textbf{Proposed Method} & \textbf{Application Domain} & \textbf{Key Features} & \textbf{Security Triad} \\
    \midrule
    Xue et al., 2022~\cite{xue2022differential} & Acies: Privacy-preserving classification framework for IoT devices & Multi edge computing applications & - Perturbs the feature extraction process instead of input data \newline - Minimizes impact on classification accuracy & Confidentiality, Integrity \\
    
    Huang et al., 2021~\cite{huang2021privately} & Personalized sampling and aggregation mechanisms with the Laplace process & IoT Data Publication & - Assigns unique sampling probabilities to each data record \newline - Improves accuracy in data aggregation & Confidentiality, Integrity \\
    
    Chong \& Malip, 2024~\cite{chong2024may} & Correlated differential privacy for spatial-temporal and user correlations in IoT & Location specific applications & - Handles spatial-temporal correlation and adversary prior knowledge \newline - Provides formal privacy guarantees & Confidentiality, Integrity \\
    
    Zheng et al., 2019~\cite{zheng2019differentially} & Compressed sensing-based differential privacy mechanism & High-Dimensional IoT Data Queries & - Reduces computational cost via compressed sensing \newline - Ensures differential privacy for linear query responses & Confidentiality, Integrity \\
    
    Kserawi et al., 2022~\cite{kserawi2022privacy} & Differentially-private model for billing accuracy using noise and lightweight cryptographic primitives & Smart Grids and Low-End IoT Devices & - Ensures billing accuracy using privacy-preserving noise \newline - Combines cryptographic primitives with fog computing & Confidentiality, Integrity, Availability \\
    
    \bottomrule
  \end{tabular}
\end{table}

Traditional cloud-enabled IoT frameworks are revolutionized by IoT devices that provide bidirectional network performance, such as those enabled by edge computing. Although this change decreases dependency and latency, it creates new security threats. The advent of local servers makes it impossible to transfer privacy-preserving cloud solutions to edge computing settings. Although edge computing services can be trusted, there are underlying risks when choosing a system for resource-constrained edge-based IoT environments. For example, data reconstruction attacks on less secure edge servers could be exploited for public safety or commercial objectives. Differential privacy, which anonymizes sensitive data by introducing noise, presents a potential method for protecting privacy in edge-based IoT systems. The study \cite{xue2022differential} proposed Acies, a privacy-preserving classification framework for IoT devices operating in an edge computing setting. Acies is designed to disrupt the feature extraction component, addressing privacy concerns when ML models are offloaded from the cloud to edge nodes, rather than directly perturbing the input data in the training set, which can significantly impact classification accuracy. 

Due to the widespread usage of IoT devices in our daily lives, such as wearable technology and smart home appliances, large institutions that support routine activities have access to enormous amounts of IoT data. Although DP can allay users' concerns about their personal privacy when these IoT datasets are published, its accuracy is limited. The authors \cite{huang2021privately} used personalized sampling technology to enhance the sample and aggregation methods, the Laplace mechanism, and other differentially private processes. This allowed for the private publication of IoT data sets through these mechanisms. Specifically, enhanced processes assign a unique sample probability to each data record, thereby increasing accuracy. DP can be used to address privacy issues related to the publication of aggregated data and user correlation, thereby protecting the privacy of location data released in the IoT. The study \cite{chong2024may} examined the intricate and simultaneous effects of spatial-temporal correlation, user correlation, and adversary prior knowledge on the privacy leakage of the DP mechanism. A correlated differential privacy approach was proposed to provide formal privacy assurance against the additional privacy leakage caused by these circumstances.

Additionally, in high-dimensional IoT data, the required noise in queries is proportional to the size of the data domain, which is exponential in the dimensionality, rendering the DP procedures currently in use ineffective and computationally costly. The authors \cite{zheng2019differentially} developed a compressed sensing mechanism based on the compressed sensing framework to address these issues by enforcing differential privacy while providing accurate responses to linear queries. In smart grids, masking and added noise ensure user privacy, but they may compromise billing accuracy, necessitating careful decisions. The study \cite{kserawi2022privacy} proposed a differentially private model that preserves billing accuracy while adding noise to data from a virtual charged battery. This method ensures the confidentiality and validity of data created by low-end IoT devices by combining lightweight cryptographic primitives with fog-computing data aggregation. Table \ref{tab:table3} summarizes the most recent differential privacy methods, including the proposed method, application domain, key features, and the security triad achieved.

\subsection{Consent mechanisms}

\begin{table}[htbp]
  \caption{Review of the most recent state-of-the-art consent mechanisms.}
  \label{tab:table4}
  \small
  \begin{tabular}{@{}p{1.5cm}p{3.5cm}p{5cm}p{4.2cm}@{}}
    \toprule
    \textbf{Refs} & \textbf{Proposed Method} & \textbf{Key Features} & \textbf{Benefits} \\
    \midrule

    Pinto \& Prazeres, 2024~\cite{pinto2024towards} & Fog of Things (FoT) – Personal Data Stores (PDS) paradigm & - Decentralized data management \newline - User control and consent-informed protection \newline - GDPR alignment & - Tackles privacy issues such as identification, localization, tracking, profiling, and linkage \\

    Lee et al., 2019~\cite{lee2019designing} & Asymmetric cryptography-based consent management & - Encrypted message exchange between Data Security (DS) and Data Collection (DC) \newline - Utilizes an intermediary for consent approval & - Enhances secure consent management \newline - Protects sensitive information fields \\

    Merlec et al., 2021~\cite{merlec2021smart} & Dynamic consent management via smart contracts on blockchain & - User-controlled consent \newline - Moderate resource usage \newline - High throughput, low latency & - Improves privacy policy implementation \newline - Supports GDPR compliance \\

    Alhajri et al., 2022~\cite{alhajri2022blockchain} & Dynamic consent framework for fitness data sharing using blockchain & - Focused on fitness data privacy \newline - Blockchain-enabled dynamic consent handling & - Enhances GDPR compliance \newline - Strengthens privacy in fitness data applications \\

    Rivadeneira et al., 2023~\cite{rivadeneira2023blockchain} & Consent management using permissioned blockchain & - Resolves one-time consent limitation \newline - Ensures transaction integrity \newline - Strong authentication and access controls & - Useful for auditing and compliance assurance \newline - Prevents data misuse and breaches \\

    \bottomrule
  \end{tabular}
\end{table}

The disparity between users' expressed concerns and their real behavior when utilizing IoT devices is known as the privacy paradox \cite{pathmabandu2020informed}. It is becoming increasingly challenging to obtain and provide legally acceptable consent due to the stringent criteria of the GDPR. For instance, data acquired by IoT solutions for smart buildings may be disclosed to other parties without the users' knowledge, making it difficult to distinguish between private and public information \cite{pathmabandu2023privacy}. This information might be kept in transit by outside parties. Additionally, the majority of human-centered IoT systems lack access to user-domain resources and data management tools \cite{rivadeneira2023blockchain}. Furthermore, these responsibilities are usually assigned to a central organization, necessitating a trustworthy connection, and may result in transparency issues. Hence, consent management in IoT devices is an essential procedure that takes user preferences and privacy policies into account. Table \ref{tab:table4} summarizes the most recent consent mechanisms in terms of the proposed method and key features, in addition to the benefits attained.

A user-centric informed consent paradigm is required, one that raises users' awareness of privacy invasion and data collection techniques, offers fine-grained visibility into privacy compliance, and proposes remedial measures through targeted recommendations. The Fog of Things (FoT)-Personal Data Stores (PDS) paradigm is proposed by Pinto \& Prazeres (2024) to protect personal data privacy in the IoT context \cite{pinto2024towards}. This decentralized solution empowers users with control over their data, addressing privacy issues in the FoT context. PDS offers benefits such as decentralized data management, user control, and consent-informed protection, aligning with the GDPR and other data protection regulations. This approach can help address IoT privacy threats like identification, localization, tracking, profiling, and linkage. The issue with automatic consent mechanisms is that people can see it adversely \cite{rivadeneira2023user}. According to certain models, data should be retained for consent reviews; nevertheless, this could make it more difficult to respond to urgent requests.

 Blockchain and cryptography play a significant role in the consent management of IoT devices. Lee et al. \cite{lee2019designing} proposed an asymmetric cryptography-based consent management that involves an encrypted message exchange model between a Data Security (DS) and a Data Collection (DC) through an intermediary. The DC chooses sensitive information fields and sends an encrypted message requesting consent. A dynamic consent management system architecture based on smart contracts was developed by Merlec et al. (2021) \cite{merlec2021smart}, enabling users to control their consent and allowing organizations to gather and utilize their personal information. The blockchain-enabled system used a moderate amount of resources while achieving high transaction throughput and low latencies. The study helps secure privacy and data, which could enhance the design of privacy policies. Alhajri et al. \cite{alhajri2022blockchain} addressed privacy concerns of IoT devices by proposing a dynamic consent framework based on blockchain for managing consent in fitness data sharing. Although it has limitations, such as detailed performance and scalability evaluations, the study helps secure data and privacy, which could aid in GDPR compliance and facilitate the creation of effective privacy policies for fitness data sharing. Rivadeneira et al. \cite{rivadeneira2023blockchain} proposed a model for consent management in IoT systems, offering a comprehensive solution to the issue of one-time consent. Maintaining transaction integrity through a permissioned blockchain also addresses problems with transparency. This concept, which utilizes blockchain technology instead of centralized methods, is particularly relevant to compliance assurance and auditing procedures. To protect against unwanted access, data breaches, and the misuse of sensitive information, data-sharing duties about consent management require strong authentication, access control, and key agreement procedures.

 Protecting digital privacy in IoT environments is a challenging task that requires a delicate balance of security, utility, and legal compliance. While data encryption is considered an essential measure against unauthorized access, it minimizes the damage even if data is intercepted or breached. Mitigating the impacts of AI analytics while protecting sensitive information is achieved through differential privacy, which prevents inference from aggregated datasets. Additionally, consent mechanisms empower users to control the sharing or collection of their personal data. Nonetheless, the fast-paced evolution of IoT ecosystems makes the efficient application of these strategies challenging, requiring further innovation in lightweight encryption, adaptive privacy-preserving mechanisms, and more open consent systems that elicit greater trust from users.

 \section{Role of AI in IoT Digital Privacy} \label{section 5}

 \begin{figure}[!t]
\centering
\includegraphics[width=12cm]{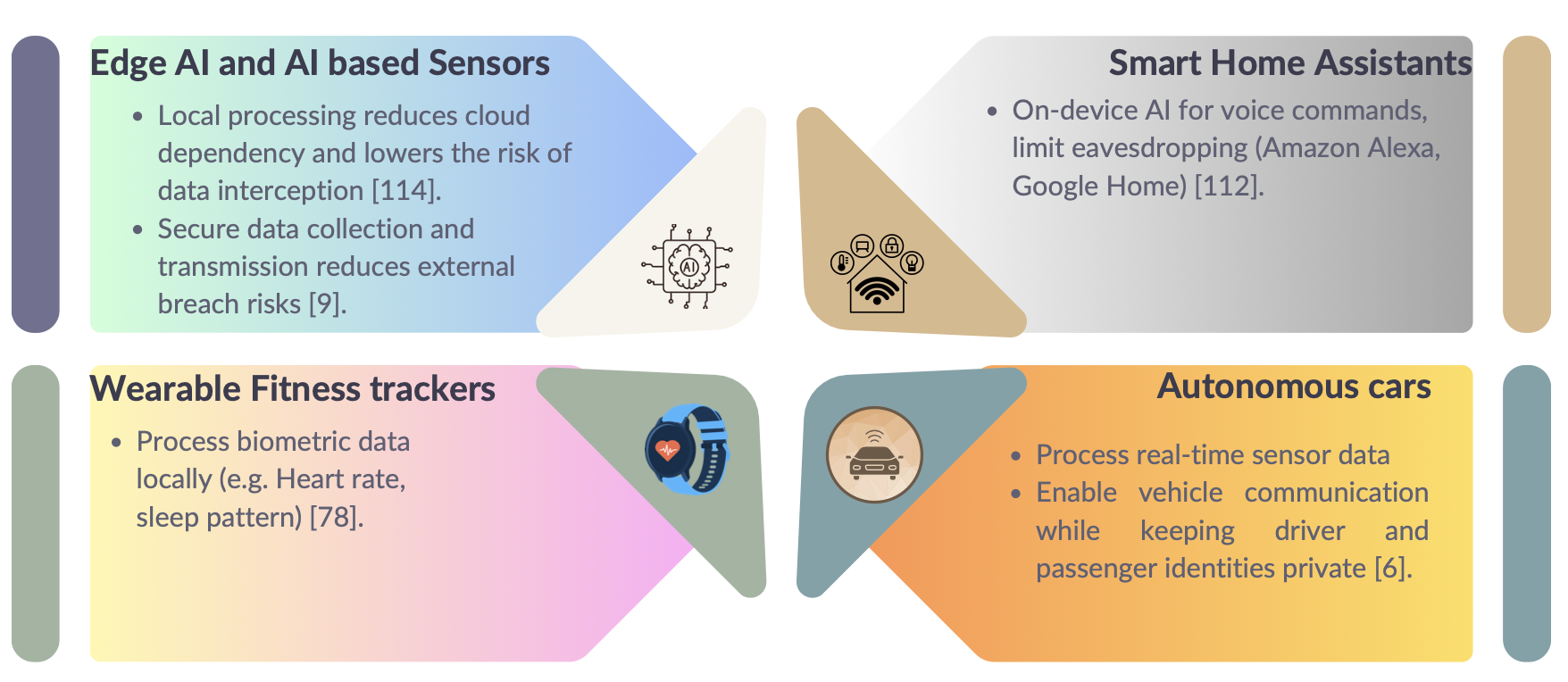} 
\caption{AIoT applications with privacy-enhancing features} 
\label{fig5}
\end{figure}

In recent years, we have witnessed a paradigm shift in the IoT ecosystem leveraging AI. For instance, AI-powered IoT systems identify trends in patient data within IoT healthcare ecosystems, thereby enhancing resource allocation, traffic control, and urban planning in smart cities \cite{marengo2024navigating}. The integration of AI and IoT is enhancing data analysis and strategic decision-making, resulting in significant investment and market expansion. By 2032, the global market for AI-powered IoT is expected to grow to \$91.7 billion, with an estimated \$3.9 trillion to \$11.1 trillion in economic effect annually by 2025 \cite{ai_in_iot_market}.

 ML is important for IoT applications since it makes predictive analytics possible and provides significant insights that potentially revolutionize these applications. As computing and communication technologies progress, ML makes it possible to analyze vast amounts of data, including those generated by IoT systems, and use the knowledge that is extracted—such as trained models—to support real-time decision-making in difficult scenarios \cite{arachchige2020trustworthy}. In contrast to other ML domains, DL exhibits high precision in speech recognition, natural language processing, and image classification applications. However, ML/DL algorithms were trained using the central location of data, which may result in increased latency, security vulnerabilities, privacy risks, and a single point of failure \cite{singh2022framework}. Differential privacy approaches \cite{xue2022differential}, Reinforcement Learning (RL) \cite{shen2022distributed}, and Federated Learning (FL) \cite{zheng2022exploring} are employed to prevent the disclosure of private data from IoT devices. These techniques improve privacy in IoT applications by allowing models to be trained on decentralized data sources while maintaining sensitive information.

 Artificial Intelligence of Things (AIoT) integrates AI-driven
knowledge of IoT devices, and this improves scalability, resilience, and efficiency \cite{parihar2023smart}. Edge AI, also known as Edge Intelligence, improves privacy by reducing dependence on central cloud servers for computing power and processing data closer to its source. AIoT analyzes data in real time without compromising user-sensitive information with external networks. AIoT is also motivated by edge computing, which moves cloud-based server data processing to the network edge for low-latency, privacy-friendly applications \cite{walia2023ai}. AIoT extends its benefits through edge computing by providing privacy features that reduce the risks of data breaches, such as unauthorized access, data leakage, and compliance issues. One example is AI-driven anomaly detection, where real-time identification of privacy threats does not wait for potential vulnerabilities to be addressed. AI-based sensors are effective in enhancing privacy, as they enable secure data collection and transfer \cite{agarwal2024artificial}. AI-based vehicular ad hoc networks (VANETs) routers protect location information while improving traffic management within Intelligent Transportation Systems (ITS). In a similar fashion, smart applications in energy, transportation, and irrigation systems capitalize on AIoT’s capacity to mask user data without compromising on functionality \cite{parihar2023smart}. Fig. \ref{fig5} presents various AIoT applications with privacy features.

AIoT adds an additional layer of privacy to other interconnected technologies. Identity management systems with strong authentication protocols minimize the chance of unauthorized access and identity fraud, as only validated users can gain entry to sensitive systems \cite{huang2023security}. Smart wearable devices also serve as an example of AIoT, helping to preserve privacy. AI algorithms incorporated in fitness trackers, such as Fitbits, process biometric data locally rather than remotely, making them less prone to privacy breaches \cite{mirmomeni2021wearables}.

Furthermore, AI-equipped smart accessories for homes, such as security cameras and smart thermostats, filter the data they collect through video conferencing to enforce user restrictions and prevent unauthorized access to privacy and surveillance footage \cite{merabet2021intelligent}. Alexa from Amazon Echo and the Google Home app have recently incorporated on-device AI to assist in voice command processing, eliminating eavesdropping opportunities and data interception \cite{sutar2024review}. Similarly, Tesla's self-driving cars utilize AIoT to process real-time sensor data, enabling these vehicles to communicate with other vehicles and infrastructure while protecting the driver's and passengers' identities \cite{acharjeedriverless}.

With the implementation of encrypted private communication and federated AI models, AIoT enables privacy protection across various domains by decentralizing data processing. These changes further demonstrate that AIoT extends beyond the application of IoT, also enhancing security, compliance, and user access to sensitive information.

 \subsection{Challenges in Employing AI for Digital Privacy in IoT}

 \begin{figure*}[!t]
\centering
\includegraphics[width=5in]{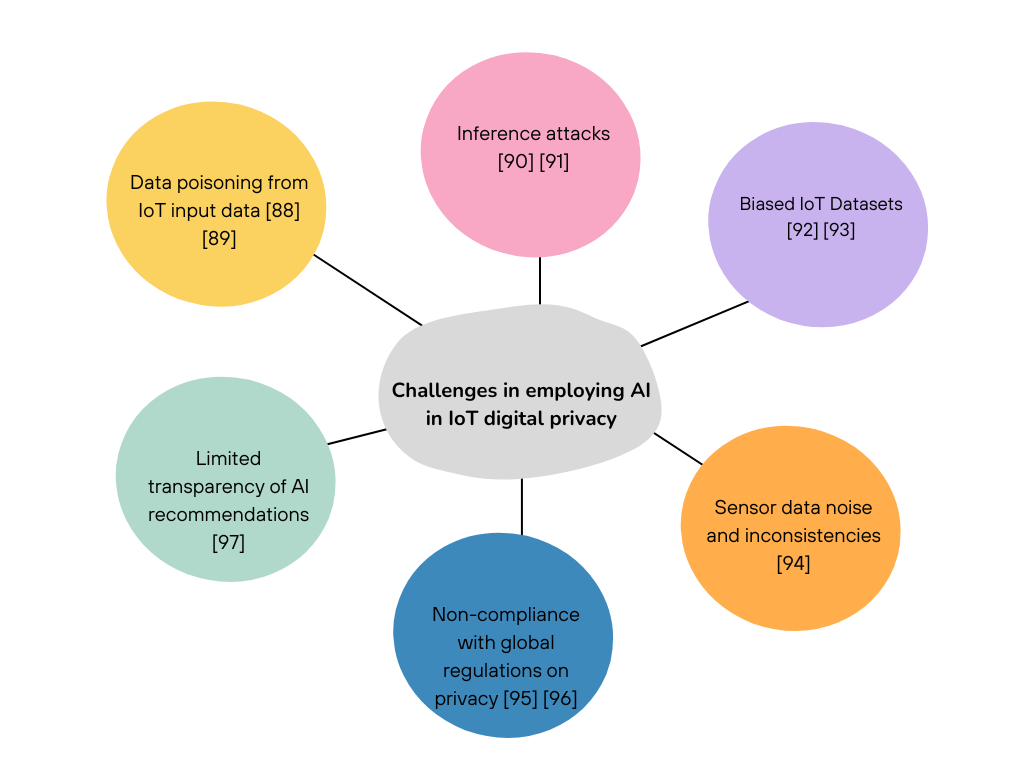}
\caption{Challenges in employing AI for Digital Privacy in IoT} 
\label{fig6}
\end{figure*}

 The convergence of AI and IoT presents a significant challenge in balancing the power of data analysis with privacy protection. The interconnected nature of IoT devices magnifies the impact of privacy breaches, emphasizing the need for robust measures to mitigate risks. The following lists the main challenges to using AI for digital privacy in the IoT  (see Fig. \ref{fig6}).

 \begin{itemize}
     \item Data poisoning attacks:- Data poisoning is a strategy where attackers compromise randomly chosen samples to increase loss \cite{geng2024privacy}. The poisoned sample is then inserted into the legitimate sample set after the attack, compromising model performance and producing inaccurate results. Attacks using data poisoning have the ability to alter private information, resulting in false model predictions and privacy violations \cite{karimyenhancing}. Digital privacy can be threatened if attackers can deduce personal information from the model. 
     \item  Inference attacks:- The statistical correlations between public and private data are exploited by inference attacks, which provide serious privacy threats \cite{wang2021membership}. Despite the lack of direct access to private data, such attacks can expose sensitive information about people or systems. Inference attacks, in which adversaries attempt to infer personal information from data, can compromise smart healthcare systems. The most common type of attack is de-anonymizing one, which aims to re-identify individuals using anonymized user data \cite{zaman2024privacy}.
     \item Biased IoT datasets: There are serious ethical issues with the rapid adoption of AI in IoT, particularly regarding bias and fairness in ML models \cite{yang2023differentially}. The massive volumes of data bring up significant privacy concerns that IoT devices collect and handle. Model parameter analysis can be used to infer sensitive user data, which could result in security breaches \cite{orlandi2023entropy}. AI-driven systems may face ethical dilemmas due to the use of biased IoT datasets, which have the potential to reinforce or exacerbate existing social biases.
     \item Sensor data noise and inconsistencies:- The widespread use of IoT devices raises privacy issues since they gather sensitive data that could be manipulated \cite{liu2024multi}. Sensor data may be affected by various factors, including malfunctioning equipment, external influences, and device interference. 
     \item Non compliance with privacy laws:- Recent developments by the integration of AI and the IoT need to abide by laws like the HIPAA, GDPR \cite{almeida2025innovations}. The regulations governing privacy vary greatly from place to place. For example, the USA places more emphasis on technology innovation than the European Union, which results in contradictions and difficulties with enforcement \cite{frahm2022fixing}. Digital user trust may be undermined by data misuse, bias, and discrimination resulting from noncompliance with these privacy regulations.
     \item Limited transparency: The interconnectedness of the IoT ecosystem may amplify the impact of AI's lack of transparency \cite{burton2024depth}. For example, IoT devices that gather patient data in the healthcare industry must guarantee the security and transparency of the AI algorithms that process this data. Privacy issues arise from the limited transparency of AI suggestions, which affect people's capacity to give their informed consent for the use of their data.
 \end{itemize}

 Privacy protection is crucial in fostering trust among users and stakeholders, requiring techniques such as encryption, anonymization, and decentralized data processing using blockchain technology. These solutions should be flexible enough to accommodate the varied and resource-constrained nature of IoT devices.

\subsection{Existing solutions in ensuring digital privacy in AIoT applications}
\subsubsection{ML/DL solutions:-}

 \begin{figure*}[!t]
\centering
\includegraphics[width=6in]{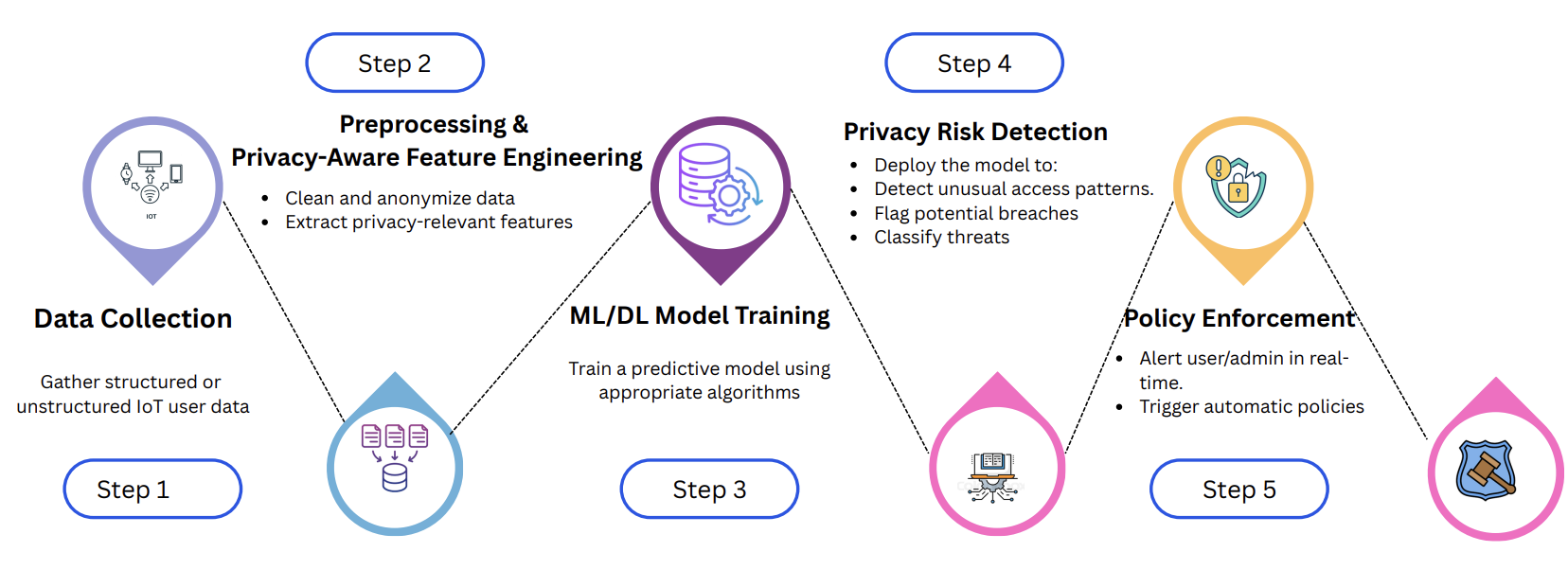}
\caption{General ML/DL framework solution.} \label{fig 3}
\label{fig7}
\end{figure*}
AI techniques in various IoT applications, such as the IIoT, cloud and edge, and fog computing paradigms, make automation and data analysis possible. ML/DL systems help security personnel spot and address privacy breaches early by analyzing user demographics, employment information, and behavior to detect suspicious behavior. A general ML/DL framework for privacy preservation is shown in Fig. \ref{fig7}. The following steps comprise a general ML/DL framework for a privacy-preserving solution: data collection, which involves gathering sensor or other IoT application-based structured or unstructured user data. After that, pre-process the data by cleaning it and performing anonymization procedures, such as deleting personal identifiers like IP addresses, names, and locations. Next, extract privacy-aware features, such as time, frequency of access, and location changes. The ML/DL model is then constructed using suitable algorithms. The model is used to identify anomalous access patterns, detect potential threats, and categorize those risks. Ultimately, alarms are sent out, and digital privacy regulations are implemented accordingly.

However, DL and other ML models still struggle to create a generalized, reliable architecture that exposes the models' contexts, semantics, and privacy breaches. Liu et al. \cite{liu2020padl} proposed a privacy-aware and asynchronous deep learning framework that enables multiple data collection sites to train deep neural networks while maintaining the confidentiality of individual private information. The layerwise importance propagation approach is used explicitly at each data-collecting location to determine the relative importance of each weight before uploading the local model to the cloud server. The weights are then adaptively perturbed to preserve the privacy of the training data. Through proxy re-encryption, Zhang et al. \cite{zhang2019deeppar} presented a unique privacy-preserving, asynchronous deep learning strategy that naturally offers dynamic update secrecy while protecting the privacy of user-sensitive information.

\begin{table}[htbp]
  \caption{Review of the most recent ML/DL solutions.}
  \label{tab:table5}
  \small
  \begin{tabular}{@{}p{2.5cm}p{3.5cm}p{5cm}p{4.2cm}@{}}
    \toprule
    \textbf{Refs} & \textbf{Proposed Method} & \textbf{Key Features} & \textbf{Benefits} \\
    \midrule

    Liu et al. (2020) \cite{liu2020padl} & Privacy-aware and asynchronous deep learning framework & 
    - Layerwise importance propagation to prioritize weights \newline 
    - Adaptive weight perturbation for privacy preservation & 
    - Maintains confidentiality of training data \newline 
    - Enables secure asynchronous training \\
    
    Zhang et al. (2019) \cite{zhang2019deeppar} & Privacy-preserving asynchronous deep learning with proxy re-encryption & 
    - Dynamic update secrecy \newline 
    - Protects sensitive user information using proxy re-encryption & 
    - Enhances privacy for dynamic updates in deep learning \newline 
    - Facilitates secure user-sensitive information handling \\
    
    Janardhana et al. (2023) \cite{janardhana2023detecting} & Deep learning-based privacy breach detection using CNN and RNN & 
    - CNN for binary and multiclass classifications \newline 
    - RNN for time-dependent privacy threat detection & 
    - Superior performance in detecting exploits, reconnaissance, and analysis \newline 
    - Effective for various IoT security scenarios \\
    
    Alsekait et al. (2024) \cite{alsekait2024privacy} & Hidden malware detection using Wide Residual Networks (WRN) & 
    - Uses ML-based WRN to detect malicious software \newline 
    - Focuses on compromised IoT device security & 
    - Enhances privacy by identifying hidden malware in IoT systems \\
    
    Luo et al. (2022) \cite{luo2022differential} & Differential privacy budget optimization technique for deep learning & 
    - Optimizes privacy budget for safeguarding deep model parameters \newline 
    - Balances generalization and privacy protection & 
    - Improves model generalization capability while preserving privacy \newline 
    - Effective for IoT deep learning applications \\
    
    \bottomrule
  \end{tabular}
\end{table}

Janardhana et al. \cite{janardhana2023detecting} utilized Convolutional Neural Networks (CNNs) and Recurrent Neural Networks (RNNs) to enhance a model for detecting privacy breaches, including exploits, reconnaissance, and analysis. CNN yields superior outcomes for all binary and multiclass classifications. Alsekait et al. \cite{alsekait2024privacy} proposed a sophisticated method for detecting hidden malware in IoT systems that uses an ML technique, a Wide Residual Network, which enhanced privacy protection by precisely detecting malicious software that compromises IoT device security. Luo et al. \cite{luo2022differential} proposed a differential privacy budget optimization technique based on DL in IoT to safeguard deep model training process parameters, improving the model's capacity for generalization while maintaining privacy protection. Table \ref{tab:table5} summarizes the most recent ML/DL solutions in terms of the proposed method, key features, and the benefits attained.

\subsubsection{Federated learning (FL) solutions}

\begin{figure*}[]
\centering
\includegraphics[width=5in]{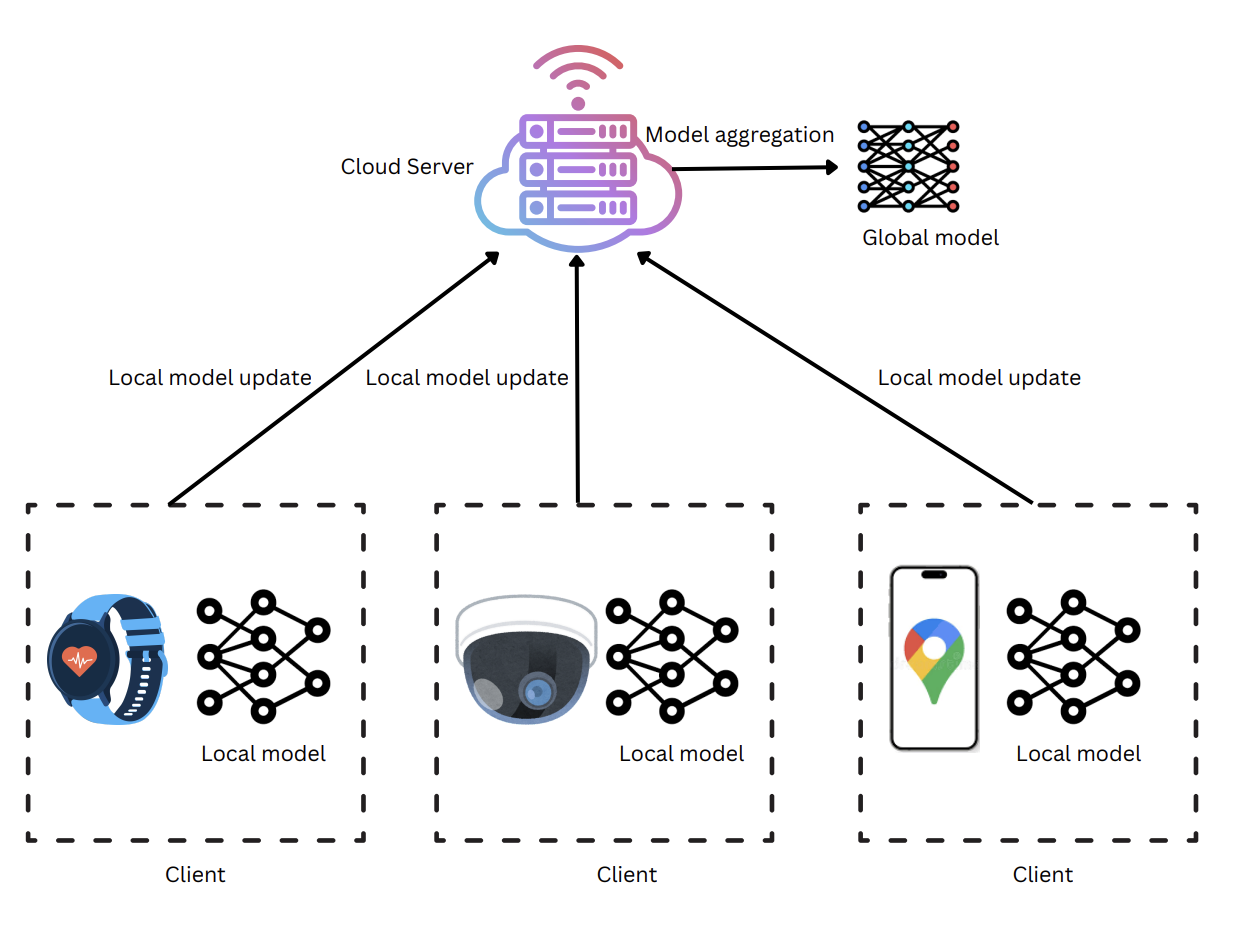}
\caption{General FL framework solution.} \label{fig8}

\end{figure*}

IoT applications require innovative AI solutions that can handle sensitive data and delay, allowing them to function locally without transferring data to a centralized organization. FL is a distributed learning approach designed to enhance privacy and minimize data delay in IoT applications \cite{aouedi2024survey}. It minimizes average loss and data sensitivity by distributing training data among devices \cite{wang2023fl4iot}. Concerns about privacy leaks and advancements in mobile technologies make FL an ideal tool for developing distributed IoT systems. It ensures that user data stays at the origin by delegating AI functions to the network edge. A general FL framework for privacy preservation is shown in Fig. \ref{fig8}. However, the complex and diverse IoT ecosystem presents security and privacy issues when using sensor data in FL \cite{rodriguez2023survey} \cite{uprety2025human}. 

\begin{table}[htbp]
  \caption{Review of the most recent FL solutions.}
  \label{tab:table6}
  \small
  \begin{tabular}{@{}p{2.5cm}p{3.5cm}p{5cm}p{4.2cm}@{}}
    \toprule
    \textbf{Refs} & \textbf{Proposed Method} & \textbf{Key Features} & \textbf{Benefits} \\
    \midrule

    Xu et al. (2024) \cite{xu2024flpm} & FL Properties Modification Scheme (FLPM) & 
    - Algorithms for property separation, selection, and control based on continuous latent variables \newline
    - Alters training data properties to prevent inference and data poisoning attacks & 
    - Enhances data preprocessing in FL \newline
    - Improves privacy by reducing inferences and protecting against poisoning attacks \\

    He et al. (2024) \cite{he2024privacy} & Privacy-enhanced FL with dual-model framework & 
    - Combines personalized and shared models for inter-client communication \newline
    - Local gradient shielding to protect training data & 
    - Protects client privacy by hiding gradients \newline
    - Reduces resource usage while maintaining data privacy \\

    Celdrán et al. (2022) \cite{celdran2022privacy} & Host-based privacy-preserving intrusion detection system (IDS) & 
    - Uses syscall fingerprints to train local and federated ML/DL models \newline
    - Identifies sensing data falsification attempts impacting IoT spectrum sensors & 
    - Enhances IoT security by detecting data falsification \newline
    - Ensures privacy during intrusion detection \\

    Wang et al. (2024) \cite{wang2024rpifl} & Reliable FL with privacy-preserving mechanisms & 
    - Lightweight network for locally hiding private data \newline
    - Multi-modal fine-grained optimization for improved model accuracy \newline
    - Adaptive differential privacy methods for dynamic privacy requirements & 
    - Protects IoT data confidentiality and privacy \newline
    - Improves FL model accuracy and adaptability in dynamic environments \\

    Zheng et al. (2024) \cite{zheng2024ppsfl} & Privacy-Preserving Split Federated Learning (PPSFL) & 
    - Combines FL and Split Learning \newline
    - Private Group Normalization layers to address data heterogeneity \newline
    - Model decomposition for easier privacy protection & 
    - Reduces the impact of data heterogeneity \newline
    - Enhances privacy protection and manageability of FL systems \\

    \bottomrule
  \end{tabular}
\end{table}

FL models can be compromised by malicious actors using IoT devices to spread poisoned information throughout time-sensitive computer systems. Hackers may also use privacy inference attacks to obtain private data from other FL devices. As a result, concerns have arisen regarding data security and privacy protection when using FL. Xu et al. \cite{xu2024flpm} suggested the FL properties modification scheme (FLPM) for data pre-processing to prevent inferences and data poisoning attacks. FLPM uses algorithms for property separation, selection, and control based on continuous latent variables to alter the properties of training data. He et al. \cite{he2024privacy} suggested privacy-enhanced FL techniques for IoT data heterogeneity by incorporating a dual-model framework for every device, combining personalized and shared models to enable inter-client communication. They also suggested a local gradient shielding technique to safeguard training data. This technique hides gradients in locally learnt customized models, avoiding data reconstruction and protecting client privacy while using the fewest resources possible. Celdran et al. \cite{celdran2022privacy} introduced a host-based, privacy-preserving intrusion detection system (IDS) that uses fingerprints based on syscalls to train local and federated ML/DL models that can identify sensing data falsification attempts that impact IoT spectrum sensors.

Wang et al. \cite{wang2024rpifl} proposed a reliable federated learning approach with privacy preservation for IoT, which consists of three essential parts. The private data is first locally hidden using a lightweight network, and the subsequent model training is then carried out using the output. Second, a multi-modal fine-grained model optimization is provided that significantly raises the IoT model's accuracy. Third, the confidentiality of IoT data is protected by implementing adaptive differential privacy methods adapted to dynamic privacy requirements. Zheng et al. \cite{zheng2024ppsfl} suggested Privacy-Preserving Split Federated Learning (PPSFL), a hybrid framework combining FL and Split learning. It reduces the negative impacts of data heterogeneity by adding private Group Normalisation layers to the network, and makes privacy protection more manageable with the correct model decomposition technique. Table \ref{tab:table6} summarizes the most recent FL solutions in terms of the proposed method, key features, and the benefits attained.

\subsubsection{Reinforcement learning (RL) solutions}

RL is adept at enhancing the security features of the IoT by incorporating dynamic security mechanisms that adapt over time and in response to user activities \cite{aouedi2024survey}. Although static methods exist, RL offers a comprehensive approach to dynamically optimizing digital privacy. This includes control and learning processes that minimize system resource consumption while providing adequate protection. Moreover, access control is refined through RL by reducing unnecessary data leakage from an IoT network. RL also empowers IoT devices to anticipate and respond to hostile attempts to compromise the system, offering adaptive real-time privacy protection \cite{hore2025deep}. 

\begin{figure*}[]
\centering
\includegraphics[width=6in]{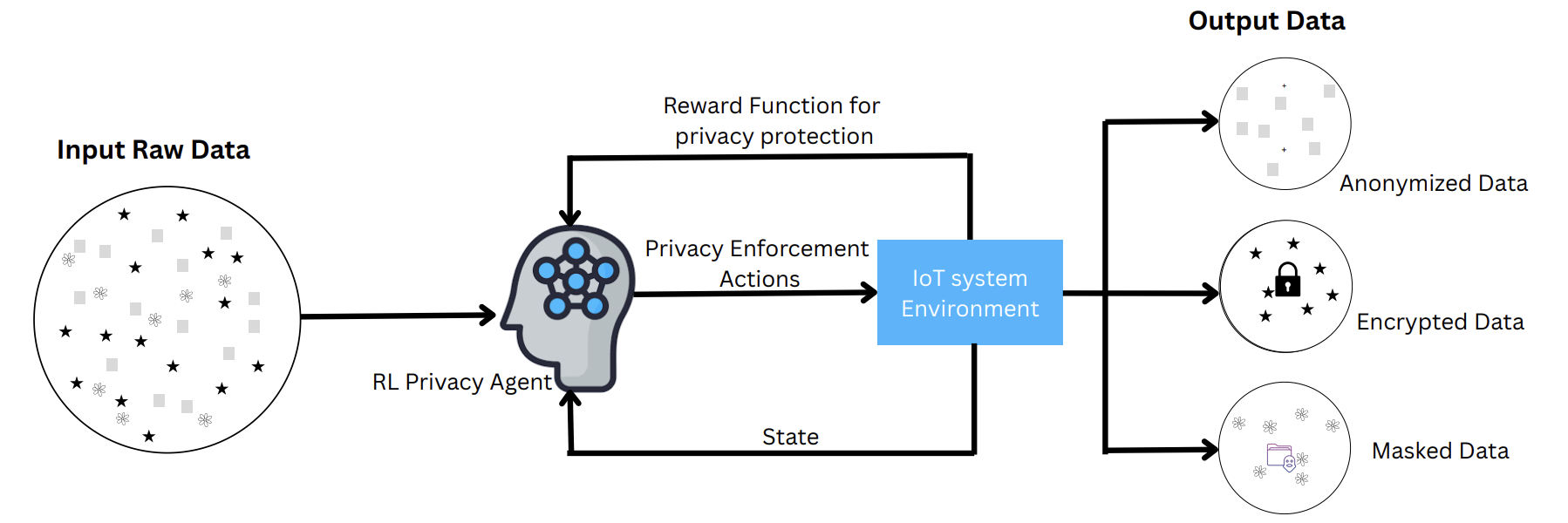}
\caption{General RL framework solution.} \label{fig9}

\end{figure*}

A general RL framework solution for privacy preservation is presented in Fig. \ref{fig9}. The overall structure begins by gathering raw input data from different IoT sensors or applications, which may include users' sensitive personal information. The RL privacy agent receives this data and decides which privacy preservation technique is most appropriate. This agent interacts with the IoT system environment, which makes decisions on anonymization, masking, and encryption. Agents learn and improve their understanding of privacy regulations over time by leveraging the reward signal that the IoT system environment provides as feedback. The output data is obtained in the form of anonymized, encrypted, or masked data according to the agent's final decision-making. This preserves the trade-off between privacy and usefulness for dynamic IoT applications.

\begin{table}[htbp]
  \caption{Review of the most recent RL solutions.}
  \label{tab:table7}
  \small
  \begin{tabular}{@{}p{2.5cm}p{3.5cm}p{5cm}p{4.2cm}@{}}
    \toprule
    \textbf{Refs} & \textbf{Proposed Method} & \textbf{Key Features} & \textbf{Benefits} \\
    \midrule

    Zhou et al. (2023) \cite{zhou2023concurrent} & Federated Reinforcement Learning (FRL) with local differential privacy & 
    - Collaboration without exchanging models; only outputs and rewards are exchanged \newline
    - Adds Gaussian noise to training gradients to satisfy local differential privacy & 
    - Protects data privacy even with untrusted servers \newline
    - Ensures secure decision-making in federated environments \\

    Alam et al. (2024) \cite{alam2024blockchain} & B-FRL: Blockchain-based Federated Reinforcement Learning & 
    - Combines blockchain with FRL \newline
    - Prevents sensor data duplication using blockchain \newline
    - Creates global FRL models to protect privacy & 
    - Ensures privacy in IoT networks \newline
    - Enhances data reliability and model integrity in IoT \\

    Baccour et al. (2021) \cite{baccour2021rl} & RL-PDNN: Privacy-aware Distributed CNN using Reinforcement Learning & 
    - Solves optimization problems for collaborative inference \newline
    - Reduces classification latency while ensuring white-box privacy \newline
    - Learns allocation policies in real-time based on resource availability and privacy levels & 
    - Protects model structure from untrusted participants \newline
    - Optimizes resource allocation in distributed networks \\

    Ding et al. (2023) \cite{ding2023enhancing} & RLAIF: Reinforcement Learning with Blockchain and Constitutional AI (CAI) & 
    - Combines RL with blockchain and CAI \newline
    - Adaptive identity management, secure data storage, and access control \newline
    - CAI ensures compliance with regulations while prioritizing security and privacy & 
    - Learns and adapts security measures based on AI-generated feedback \newline
    - Ensures compliance with privacy laws and regulations \\

    Liu et al. (2020) \cite{liu2020incentive} & Privacy-preserving incentive system using Q-learning & 
    - Solves dynamic games using Q-learning \newline
    - Double deep Q-network with dueling architecture for better strategy acquisition \newline
    - Optimizes participant benefits and data availability & 
    - Protects participant privacy in dynamic environments \newline
    - Incentivizes secure participation in collaborative platforms \\

    \bottomrule
  \end{tabular}
\end{table}

RL has been utilized to address various IoT applications, including network resource optimization, recommendation systems, robotics, and autonomous vehicles. It entails investigating the surroundings, selecting the best strategies, and adapting to changes. In an IoT ecosystem, RL provides a framework for multi-agent systems that enable devices to collaborate, learn, and perform tasks more effectively. Additionally, it is essential to edge computing, where it optimizes data management and processing activities. The edge hosts must send their local parameters to a central server using a standard reinforcement learning method. Nevertheless, this procedure raises privacy concerns because some edge hosts' processed data is likely very sensitive. Zhou et al. \cite{zhou2023concurrent} presented the idea of integration to federated reinforcement learning (FRL), in which the server and agents collaborate to make decisions without exchanging models; just the outputs and rewards are exchanged. This approach effectively protects data privacy even when an untrusted server is present by adding Gaussian noise to the training gradients of edge hosts that satisfies local differential privacy.

Alam et al. \cite{alam2024blockchain} proposed the B-FRL approach, which integrates blockchain with FRL, for safeguarding privacy in IoT networks. Blockchain technology ensures the accuracy of information by preventing the duplication of sensor data. IoT devices raise privacy concerns since they gather and use personal data. FRL creates global models, protects privacy, and handles agent problems. Baccour et al. \cite{baccour2021rl} defined collaborative inference with respect to privacy as an optimization problem that seeks to reduce classification latency while maintaining the necessary white-box privacy by shielding the model structure from untrusted participants. A novel method called RL-PDNN, based on RL, is suggested for privacy-aware distributed CNN networks to solve the optimization problem. This method learns the allocation policy and acts in real-time according to the available resources, the dynamics of the requests, and the necessary level of privacy. Ding et al. \cite{ding2023enhancing} suggested RLAIF, an innovative IoT security and privacy solution that combines blockchain technology with constitutional AI (CAI) and RL algorithms. This solution enables adaptive identity management, secure data storage, and access control within IoT networks. CAI guarantees that the system complies with the rules and regulations outlined in the constitution, prioritizing user security and privacy. At the same time, RLAIF enables the system to learn from AI-generated feedback signals and adjust its security measures accordingly. Liu et al. \cite{liu2020incentive} developed an incentive system that aimed to protect participants' privacy, ensure data availability, and optimize platforms and participants' benefits. This system used Q-learning to solve dynamic games and a double deep Q network with dueling architecture for better strategy acquisition.  Table \ref{tab:table7} summarizes the most recent RL solutions in terms of the proposed method, key features, and the benefits attained.

Integrating AI into IoT multi-layered systems indeed creates complex issues surrounding digital privacy while providing possible solutions. Mechanisms for driving privacy face challenges such as adversarial attacks, biased decisions, data leakage, and the high computational load of privacy-preserving algorithms. Nevertheless, the risk-mitigating approaches like ML/DL, FL, and RL are highly promising. The robustness of methods that guard against adversarial AI, provide fairness in automated decision-making, and balance privacy, security, and efficiency in the context of AIoT requirements are all areas that require further research; however, these advancements remain encouraging.

\subsection{Role of Privacy-Enhancing Technologies (PETs) within the Proposed Risk Framework}

The proposed privacy risk framework uses various PETs to reduce the identified risk categories in IoT ecosystems—DP guards against inference and behavioral risks by adding statistical noise, which keeps individual-level data private. FL and ML address identity-related and data manipulation risks by keeping sensitive data local at edge or fog nodes. They enable distributed model training through encrypted model exchanges. Encryption methods protect the confidentiality and integrity of sensitive data, ensuring that information remains secure even in unreliable IoT networks. Blockchain technology enhances auditability and regulatory compliance by maintaining immutable consent records and facilitating transparent data transactions between IoT devices. Additionally, RL techniques aid in privacy management by adjusting security policies and resource allocation in response to evolving threats. Therefore, these PET techniques create a strong framework that enhances digital privacy protection in IoT systems.
\section{AURA-IoT—A futuristic framework for digital privacy in IoT to address AI risks} \label{section5}

We proposed AURA-IoT—A futuristic framework for digital privacy in IoT—to mitigate AI-based privacy risks. AURA-IoT stands for Autonomous Unified Risk Adaptive Architecture for IoT privacy. Autonomous refers to the ability to handle privacy concerns independently, without direct human supervision. The term "unified" refers to combining several privacy protection techniques to stop privacy violations. The term "risk adaptive" refers to the ability to adjust to changing user choices and target dynamic digital privacy concerns. This futuristic framework integrates solutions for AI compliance, dynamic consent, explainability, adversarial robustness, transparency, fairness, and policy enforcement into a multi-layered architecture for IoT applications.

\subsection{AI compliance}
AI compliance refers to the measures and procedures that enable businesses to adhere to the regulations and laws governing the use of AI technology \cite{deshpande2024regulatory}. It is accomplished by utilizing three essential elements: accountability, regulatory alignment, and proper documentation. 

AI accountability refers to the idea that AI should be created, implemented, and used in a way that allows liable parties in an IoT environment to be held accountable for unfavorable results \cite{raja2023ai}.
Every stakeholder involved in the development of an AI module in AURA-IoT is responsible for considering the system's global impact at every stage. The framework facilitates audit trails that document the accountability of individuals and institutions responsible for creating, developing, and implementing AI systems. This ensures trust among different stakeholders in automated decision-making involving personal IoT data (e.g., smart home assistants and wearable health devices).

Regulatory alignment guarantees that current privacy-preserving practices comply with international digital privacy rules and regulations, including PIPEDA, CCPA, and GDPR \cite{guha2024ai}. AURA-IoT ensures the enforcement of digital privacy policies by utilizing dynamic consent mechanisms, user rights preservation strategies, and data minimization techniques in AI system workflows. Additionally, the model behavior complies with current privacy rules, particularly for global IoT healthcare data flow applications that allow patient data to be accessed from a different country without violating privacy laws.
\begin{figure*}[]
\centering
\includegraphics[width=5in]{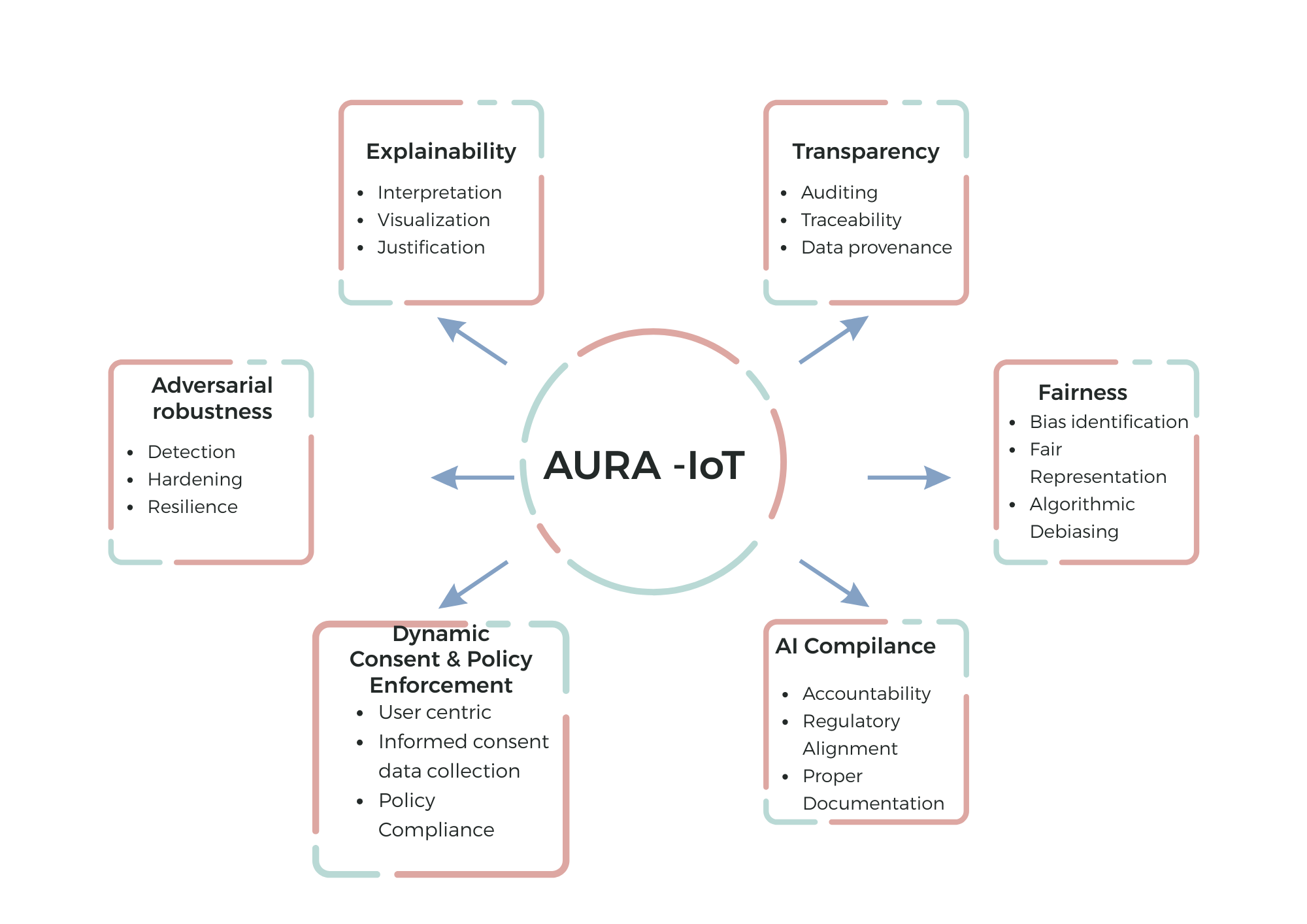}
\caption{AURA-IoT futuristic framework } \label{fig 4}

\end{figure*}

Proper documentation is conducted in AURA-IoT using information analysis, dashboards, reports, and visualizations to gain a thorough understanding of AI models' performance \cite{radanliev2024ai}. Structured reporting of these lessons will help build the capacity to combat future privacy breach events. AI/ML-driven choices are easier for non-experts to understand when dashboards that display important risk metrics, such as the possibility of exploitability, device dependencies, and the effect of failures, are included. These dashboards should provide graphical depictions of the connections between various IoT devices, risk ratings for specific systems or devices, and describe how modifications to dependencies affect the overall risk of the system.
\subsection{Dynamic consent \& policy enforcement}

The term "dynamic consent management" refers to the automated procedures and tools that businesses use to efficiently manage user permissions in IoT applications \cite{andreotta2022ai}. Dynamic consent management solutions enable ongoing, fine-grained control over user data collection, storage, and use, in contrast to static consent forms, which are often one-time agreements. These days, the discussion of consent management is dominated by the user-centric concept. Organizations must prioritize transparency and individual control in light of laws such as the GDPR and shifting customer expectations. The power dynamics between businesses and consumers are radically changed by this change, which gives more weight to meaningful informed consent experiences that foster trust and innovation. Proper consent management reduces legal risks, promotes consumer trust, and guarantees compliance with these rules. Businesses looking to successfully comply with the CCPA and GDPR must automate dynamic consent management. Organizations can enhance operational efficiency, comply with data privacy requirements, and foster customer trust in a digital environment that is becoming increasingly regulated by incorporating automated technologies.

\subsection{Explainability}
Explainability in the AURA-IoT architecture enables trust and accountability in IoT system decision-making processes by addressing the growing requirement for transparency and interpretability in AI systems \cite{hudon2021explainable}. It provides IoT users and regulators with a structured approach for understanding, evaluating, and justifying AI-driven decision-making.
This can be achieved through interpretation, visualization, and justification.  An interpretable model provides insight into the relationships between inputs and outputs, operating transparently. Interpretation aids in identifying biases or errors in vulnerable IoT applications, such as smart home assistants or healthcare monitoring. Visualizations play a crucial role in presenting decisions in a clear, human-understandable format, such as charts or dashboards, helping users understand why a particular decision is made or why a specific privacy preservation method is chosen. Finally, justification refers to the ability of AI decisions to be justified and explained to the end user, which is a crucial prerequisite for the ethical deployment of AI and regulatory compliance. It can boost user trust and improve decision-making when AI decisions are explained in a clear and understandable way.

\subsection{Adversarial robustness}
In the AURA-IoT framework, adversarial robustness focuses on protecting AI models deployed in IoT settings from targeted manipulations and sophisticated cyber threats \cite{shayea2025strategies}. This component focuses on enhancing the system's ability to withstand, recognize, and recover from attacks through methods such as detection, hardening, and resilience.

Detection involves the continuous observation of an AI's inputs and behavior to identify anomalous activities, such as adversarial spoofed sensor data and inputs that can lead the model to erroneous conclusions. Hardening also encompasses proactive measures, such as adversarial training, noise injection, or input validation, designed to fortify the model and defend it against exploitation. Lastly, resilience ensures that the system can sustain a certain level of performance acceptability and restore operations swiftly when an ongoing active attack or data corruption occurs.

\subsection{Transparency}

In the AURA-IoT framework, transparency refers to understanding how AI makes decisions in IoT settings, why it produces specific results, and what data it is using \cite{felzmann2020towards}. This can be done through auditing, traceability, and data provenance.

As an IoT transparency mechanism, auditing focuses on the retention and examination of system logs, which document all interactions with the system, post-system data accessing, processing, and sharing. This offers privacy practices in safeguarding sensitive information and revealing anomalies or misuse of the system. Traceability is the capacity to monitor and record information and decisions made by an AI system over the course of its lifetime.
This enables linking results to the inputs, models, and processes that produced those results. This is vital for appreciating and validating AI decisions in real-time, especially in AI-driven technologies. Data provenance refers to having a complete and reliable history that includes the origin of the data, the methods used in data collection, and the data processing methods employed. Data provenance focuses on issues of reliability, accountability, and integrity, even when there is a breach of data or misuse of data or information.

\subsection{Fairness}
In the AURA-IoT framework, fairness in decisions generated by AI systems is achieved by treating individuals with risk equally, regardless of their demographics, behavior, or context, ensuring equal predictive performance measures across groups, independent outcomes, and not explicitly using protected attributes \cite{zhang2024ai}. Fairness can be achieved through bias identification, fair representation, and algorithmic debiasing. 

Bias identification is the process of uncovering hidden or systemic biases in training data or AI outputs that may disproportionately affect specific groups. This involves being vigilant for unfair treatment in decision-making outcomes or skewed data distributions. To reduce the possibility of discrimination, fair representation ensures that protected characteristics such as gender, age, ethnicity, or disability are appropriately anonymized in datasets. Techniques that balance data representation across groups might also be used. By reducing bias through methods such as reweighting data, fairness-aware loss functions, or post-processing corrections, algorithmic debiasing involves modifying model training or decision logic to ensure equitable results.

The AURA-IoT framework offers a promising, futuristic framework to address AI-related privacy risks in the IoT environment. It combines the key components such as AI compliance, dynamic consent, explainability, adversarial robustness, transparency, fairness, and policy enforcement to ensure privacy, utility, trust, and regulatory compliance, with AI decision-driven systems in a dynamic IoT environment.

\subsection{AURA-IoT System Design and Workflow}

\begin{figure}[]
\centering
\includegraphics[width=4in]{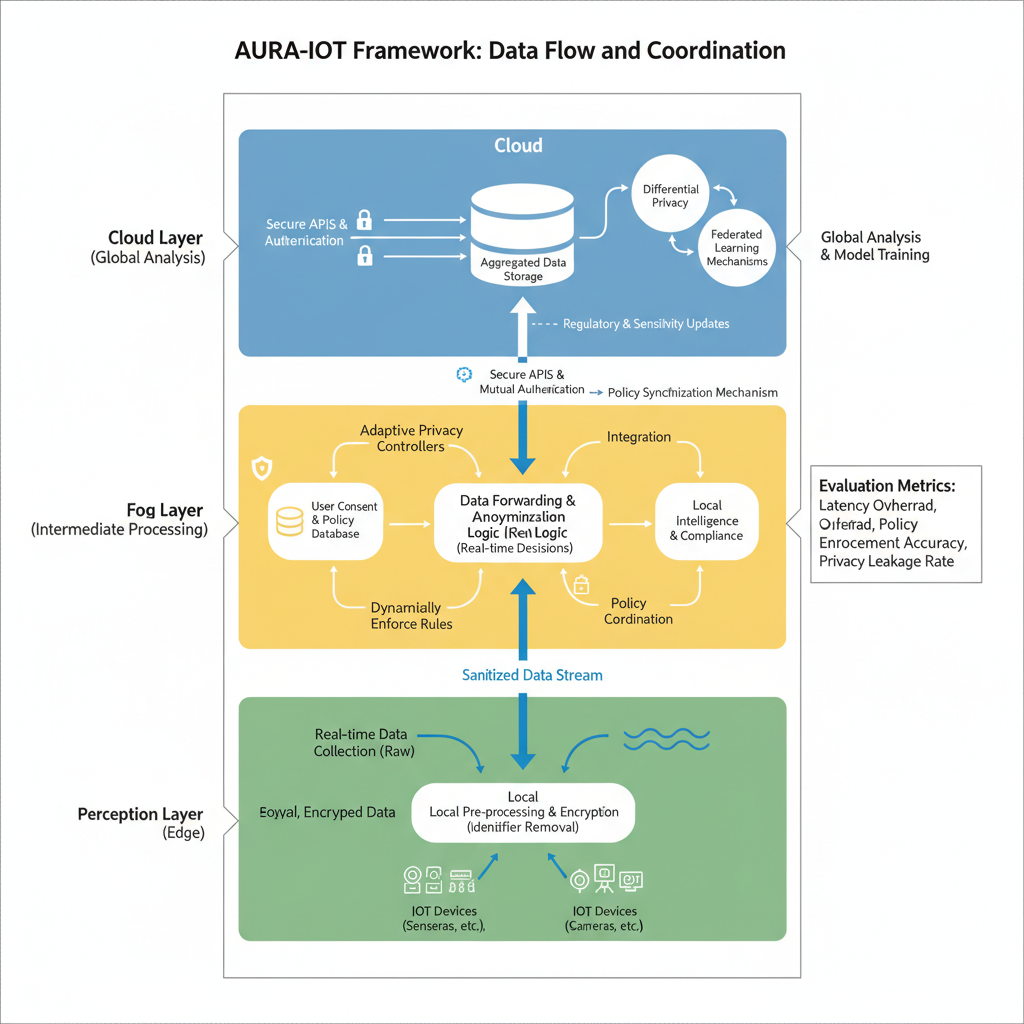}
\caption{AURA-IoT system workflow  } \label{fig 5}
\end{figure}

To enhance the technical clarity and applicability of the AURA-IoT framework, this subsection outlines its data flow, coordination between layers, and component-level functionality. The framework features a hierarchical, multi-layered architecture comprising perception, fog, and cloud layers, each with clearly defined privacy-preserving roles. In the perception layer, IoT devices gather real-time data from the environment.  The input data is locally processed to eliminate identifiers, and lightweight encryption is applied. This encrypted data is sent to the fog layer, where policy coordination is performed by dynamic privacy controllers who enforce sensitive data sharing policies considering user consent, risk levels, and compliance requirements.  The fog layer is an intermediary, combining local intelligence with compliance logic. This allows for real-time decisions on data forwarding or anonymization. The cloud layer aggregates data for global analysis and model training, utilizing differential privacy and federated learning to ensure confidentiality and fairness. Data flows between layers through secure APIs and mutual authentication, guaranteeing traceability and auditability at each transition. Coordination between these layers is handled by a policy synchronization mechanism that adjusts privacy rules in response to updates to user consent, data sensitivity, or regulatory changes. Evaluation of framework components can be based on metrics like latency overhead, policy enforcement accuracy, and privacy leakage rate. By detailing these operational aspects, the AURA-IoT framework builds a technically verifiable and flexible structure, supporting privacy-aware decision-making across IoT systems. Fig.  \ref{fig 5} presents the AURA-IoT system data workflow.

\subsection{Integration of Privacy-Enhancing Technologies within AURA-IoT}
To enhance the completeness and relevance of the AURA-IoT framework, this section directly maps PETs to the identified digital privacy risk categories. Each technology is integrated into specific layers of the AURA-IoT architecture to mitigate risks and enhance privacy assurance. Data encryption and secure key management primarily address identity-related and data manipulation risks at the perception and edge layers by ensuring confidentiality and preventing unauthorized access or changes. DP methods reduce inference and behavioral risks within the cloud and AI layers by limiting the chance of reidentifying individual data during model training and aggregation. FL further enhances this protection by storing raw data locally while allowing for shared intelligence. Dynamic consent methods and automated policy enforcement tools mitigate regulatory risks by ensuring user control and compliance with legal requirements, such as the GDPR and HIPAA. By integrating these privacy-protecting methods with the AURA-IoT risk taxonomy, the framework transitions from a concept to a practical privacy architecture that can address real-world AIoT vulnerabilities.

While AURA-IoT builds on basic AI ideas, it significantly differs from standard AI solutions by incorporating privacy-preserving tools and IoT-specific operational layers. Traditional AI frameworks primarily focus on enhancing model performance and accuracy, often overlooking the importance of multi-source data privacy, regulatory compliance, and consent-driven interactions. In contrast, AURA-IoT presents a risk-aware, layered AI structure that adjusts privacy and policy enforcement based on data sensitivity, user consent, and contextual risk levels. The framework incorporates FL and RL components in both fog and cloud layers to facilitate distributed intelligence and flexible decision-making, thereby reducing exposure to raw data. Additionally, DP, encryption, and blockchain-aided audit methods ensure that every AI action remains accountable, transparent, and compliant with evolving privacy regulations. These design improvements make AURA-IoT a privacy-smart AI architecture specially designed for IoT ecosystems, rather than just an extension of standard AI. By combining clarity, dynamic consent management, and policy synchronization across layers, AURA-IoT surpasses the typical limitations of AI to address the key issues of privacy, transparency, and compliance in diverse IoT environments. Table 9 maps PETs to AURA-IoT and risk categories.

\begin{table}[h!]
\centering
\caption{Mapping of Privacy-Enhancing Technologies to AURA-IoT Risk Categories}
\renewcommand{\arraystretch}{1.2}
\setlength{\tabcolsep}{4pt}
\begin{tabularx}{\columnwidth}{l l X}
\hline
\textbf{Technology} & \textbf{Mitigated Risk Category} & \textbf{Integration Point within AURA-IoT} \\
\hline
Data Encryption & Identity-oriented, Data Manipulation & Applied at the perception and edge layers to secure sensor data and prevent unauthorized access or alteration. \\
Differential Privacy & Inference, Behavioral & Implemented at the cloud/AI layer to prevent reidentification and limit sensitive data inference in model training. \\
Federated Learning & Behavioral, Inference & Deployed at fog and cloud layers to allow collaborative learning without raw data sharing. \\
Adversarial Robustness Techniques & Data Manipulation & Embedded in AI modules to resist malicious input manipulation and biased model behavior. \\
Dynamic Consent Mechanisms & Regulatory, Behavioral & Integrated across user and regulatory layers to ensure user-centric, real-time data access control. \\
Policy Enforcement Engines & Regulatory & Enforced at all layers to maintain compliance with evolving data protection standards (GDPR, HIPAA). \\
\hline
\end{tabularx}
\label{tab:pet_mapping}
\end{table}

\section{Future scope of Digital Privacy in IoT}\label{section6}
This section discusses the future of digital privacy in the IoT, which is expected to gain more interest in the subsequent year. Future IoT implications for digital privacy present both potential and challenges, necessitating a proactive response to security and privacy issues.

\subsection{Post-quantum cryptographic
authentication schemes}

The emergence of quantum computing could make conventional cryptography techniques susceptible. Research is underway to protect IoT systems from potential quantum threats using post-quantum cryptography (PQC). PQC includes lattice-based, code-based, and multivariate cryptography algorithms that are immune to the sophisticated computing power of quantum computers \cite{mansoor2024pqcaie}. PQC is essential for protecting IoT data from quantum computing attacks and ensuring user data security and privacy in this constantly evolving digital age.

\subsection{Privacy-by-design (PbD) schemes}

Developers can take data privacy into account during the design phase of IoT development by utilising privacy-by-design (PbD) techniques \cite{alhirabi2024designing}. The capabilities of an effective PbD tool should include (i) helping developers fulfil privacy needs in less regulated sectors and (ii) educating them about privacy while they use the tool.

\subsection{Integration with Innovative Technologies}
A breakthrough in privacy protection is being heralded by generative AI, which uses AI algorithms to create synthetic data that resembles genuine training data \cite{hazra2024review}. As AI technology develops and improves, it might make gadgets more intelligent in gathering and analyzing data while protecting user privacy. The landscape of privacy issues is also expected to evolve due to the growing sophistication of technologies like edge computing, which might efficiently divide data processing and save bandwidth \cite{abd2020controlled}. Edge computing can enhance IoT security architecture by reducing the strain on networks and the volume of data transmitted over them, making it more challenging for potential attackers to compromise privacy. 
To summarize, the integration of AI, FL, Blockchain, RL, and differential privacy in the IoT may offer several advantages, including enhanced security, improved incentive systems, enhanced privacy protection, and accelerated model training. However, for integration to be successful, issues including resource limitations, security worries, privacy risks, and ethical concerns about fairness, transparency, and digital ethics must be resolved.
\subsection{Lightweight Encryption Techniques}
The need for lightweight cryptographic solutions is growing in embedded applications of smart devices connected to IoT because of their restricted size, computing power, and battery life \cite{singh2024advanced}. NIST defines lightweight cryptography as a branch of cryptography that addresses the needs of rapidly expanding applications that frequently utilize intelligent, low-power devices. In February 2023, NIST chose the Ascon family of algorithms as the standard for lightweight cryptography due to this family’s powerful efficiency and robust security features \cite{sonmez2024ascon}. The Ascon suite features recognized authenticated encryption and hashing algorithms that are optimized for high performance in constrained devices. This decision was made after a multi-year assessment process in which NIST evaluated multiple candidates for budgetary constraints, security, performance, and flexibility. The expectation of adopting Ascon is to improve IoT application data protection while retaining security and efficiency. Besides Ascon, NIST’s Lightweight Cryptography Project has also investigated the logic of IoT-enabled block ciphers hashing and authenticating for constrained devices such as PRESENT and SPECK and lightweight authentication methods \cite{NIST-SP-800-232}. These changes enable the protection of IoT ecosystems, which is critical in areas that are most vulnerable and where traditional cryptography methods would be too expensive in terms of processing resources.

\subsection{Regulatory Compliance}
Strong, standardized, and automated compliance systems are essential for addressing IoT privacy issues and ensuring user trust and data security. Concerns regarding data breaches, personal privacy, and the ethical use of gathered data have increased significantly with the widespread adoption of IoT devices across various industries.  Since GDPR and HIPAA represent major changes over the last 20 years, data privacy is a major compliance concern in DFL. Organisations must abide by GDPR and HIPAA to guarantee the security of the personal data of EU individuals. Methods such as Deep Federated Learning can help preserve regulatory compliance and protect data privacy \cite{abbas2023exploring}.

The future of digital privacy in the IoT will be shaped by advancements in encryption, the adoption of privacy by design principles, and the integration of new technologies. While post-quantum cryptography will shield data from quantum attacks, privacy-by-design principles will integrate security from the outset of IoT development. Lightweight encryption approaches will protect privacy on IoT devices with low resources. As regulations like GDPR and HIPAA continue to evolve, organisations must implement robust compliance frameworks to protect user data. Addressing security risks, transparency, and ethics concerns will be necessary for a strong, privacy-preserving IoT ecosystem.

\section{Conclusion} \label{con}

In this survey, we have examined digital privacy in the IoT, addressing significant challenges, existing solutions, and the role of AI in transforming privacy protection. The proposed framework for classifying digital privacy risks offers a systematic approach to understanding and mitigating privacy risks in IoT systems. This framework categorizes digital privacy risks into five key areas: identity-oriented risks, behavioral risks, inference risks, data manipulation risks, and regulatory risks.

Current IoT digital privacy solutions are critically analyzed, emphasizing advances in consent mechanisms, access control, and encryption, while highlighting areas that require further study. The discussion of AI's revolutionary role in IoT privacy highlights its uses in intelligent data management, anomaly detection, and real-time threat detection. However, the incorporation of AI also presents problems that need to be carefully considered, such as biases, transparency concerns, and resource limitations.

This paper also presents the AURA-IoT framework, a promising prospective framework for addressing privacy risks associated with AI in the context of IoT. It integrates the essential elements—AI compliance, explainability, adversarial robustness, dynamic consent, transparency, fairness, and policy enforcement—to guarantee privacy, utility, trust, and regulatory compliance for AI decision-driven systems in a dynamic IoT setting.

The study also highlights several interesting areas for future research in IoT digital privacy, including the development of AI-driven privacy frameworks, flexible solutions for ever-changing IoT environments, and the standardization of international privacy policies. Future research should focus on integrating innovative technologies, such as blockchain, federated learning, and quantum computing, with IoT privacy solutions.

This study aims to assist researchers, developers, and regulators in enhancing IoT digital privacy by providing a systematic framework and exploring innovative approaches. The study enhances active controls over personally identifiable information by IoT digital users, reinforces data protection within IoT ecosystems, and boosts users’ trust in privacy-sensitive IoT systems. It also aims to improve the role of AI in detecting and identifying network threats’ anomalies and automating the enforcement of privacy policy provisions.

\bibliographystyle{unsrt}

\bibliography{cas-ref}




\end{document}